\documentclass[manuscript]{aastex} 
\usepackage{lineno}
\usepackage{amsmath}
\usepackage{amssymb}
\usepackage{natbib}
\usepackage{graphicx}
\usepackage{dcolumn}
\usepackage{bm}
\usepackage{multirow}

\shorttitle{PARALLEL VS. OBLIQUE ALFV\'EN-CYCLOTRON WAVES}
\shortauthors{Maneva et al.}

\begin{document}

\title{Dissipation of parallel and oblique Alfv\'en-cyclotron waves -- implications for 
minor ion heating in the solar wind}

\author{Y. G. Maneva}
\affil{Centre for mathematical Plasma Astrophysics, KU Leuven, B-3001 Leuven, Belgium}

\email{yana.maneva@wis.kuleuven.be}

\author{Adolfo F. Vi$\tilde{\mathrm{n}}$as}
\affil{NASA Goddard Space Flight Center, Greenbelt, MD 20771, USA}

\author{Pablo S. Moya}
\affil{Department of Physics, Catholic University of America, Washington DC 20064, USA}
\affil{NASA Goddard Space Flight Center, Greenbelt, MD 20771, USA}

\author{Robert T. Wicks}
\affil{NASA Goddard Space Flight Center, Greenbelt, MD 20771, USA}


\author{Stefaan Poedts}
\affil{Centre for mathematical Plasma Astrophysics, KU Leuven, B-3001 Leuven, Belgium}




\begin{abstract}

We perform 2.5D hybrid simulations with massless fluid electrons and kinetic particle-in-cell ions to study the temporal evolution of ion temperatures, temperature anisotropies and velocity distribution functions in relation to the dissipation and turbulent evolution of a broad-band spectrum of parallel and obliquely propagating Alfv\'en-cyclotron waves. The purpose of this paper is to study the relative role of parallel versus oblique Alfv\'en-cyclotron waves in the observed heating and acceleration of minor ions in the fast solar wind. We consider collisionless homogeneous multi-species plasma, consisting of isothermal electrons, isotropic protons and a minor component of drifting $\alpha$ particles in 
a finite-$\beta$ fast stream near the Earth. The kinetic ions are modeled by initially isotropic Maxwellian velocity distribution functions, which develop non-thermal features and temperature
anisotropies when a broad-band spectrum of low-frequency non-resonant, $\omega \leq 0.34 \Omega_p$, Alfv\'en-cyclotron waves is imposed at the beginning of the simulations. The initial plasma parameter values, such as ion density, temperatures and relative drift speeds, are supplied by fast solar wind observations made by the \textit{Wind} spacecraft at 1AU. The imposed broad-band wave spectra is left-hand polarized and resembles \textit{Wind} measurements of Alfv\'enic turbulence in the solar wind. The imposed magnetic field fluctuations for all cases are within the inertial range of the solar wind turbulence and have a Kraichnan-type spectral slope $\alpha=-3/2$. We vary the propagation angle from $\theta= 0^\circ$ to $\theta=30^\circ$ and $\theta=60^\circ$, and find that the minor ion heating is most efficient for the highly-oblique waves propagating at $60^\circ$, whereas the protons exhibit perpendicular cooling at all propagation angles.  

\end{abstract}


\section{Introduction}
The dissipation of fluid-scale fluctuations in collisionless plasmas can occur via a turbulent cascade, followed by different kinds of wave-particle interactions. The partitioning of energy between minor ions, protons and electrons and the efficiency of the heating depends on the characteristics of the waves, the wave vector direction and the anisotropy of the fluctuations carrying energy at small scales. Anisotropic fluctuations here refer to different wave power in parallel and perpendicular direction with respect to the background magnetic field.
Spacecraft observations of magnetic field fluctuations in the solar wind often provide ambiguous single-point measurements, which makes it hard to uniquely determine the type of the observed fluctuations or their angle of propagation. While parallel Alfv\'en waves and Alfv\'enic turbulence are ubiquitous in the solar wind and their existence has been unambiguously proven throughout the heliosphere \citep{Bruno:05}, much less is known about obliquely propagating waves in the interplanetary medium. The most common electromagnetic fluctuations, which can interact with the particles at the ion scales are parallel and oblique ion-cyclotron (ICWs), as well as kinetic Alfv\'en waves (KAWs). Such plasma waves at kinetic scales are very hard to detect and identify with a single spacecraft as their interpretation relies on preliminary assumptions. Yet, recent analysis of solar wind data from Helios, MESSENGER, STEREO, \textit{Wind} and Cluster spacecraft show a compelling evidence for the existence of parallel and oblique ion-cyclotron waves ICWs from $0.3AU$ to $1AU$ \citep{Jian:09,Jian:10,Jian:14}, as well as KAWs \citep{Sahraoui:10,Podesta:11a,He:12a,Salem:12} near the Earth. \citet{Jian:09,Jian:10} have shown that the ICWs are rather dispersive with a wide range of wave numbers and frequencies, and statistically significant fraction of events propagate at small angles with respect to the direction of the ambient magnetic field.
Lately, reconstruction of the magnetic helicity based on Ulysses and STEREO/IMPACT/MAG data from the undisturbed solar wind suggests that the solar wind turbulence at the ion scales can be reconstructed by a superposition of parallel and oblique ICWs/KAWs \citep{Podesta:11a,He:12b}. What role each of the different types of waves plays in relation to wave-particle interactions, anisotropic heating and differential acceleration is still an open question, which remains to be solved.
The purpose of this paper is to compare the ion heating by parallel and oblique Alfv\'en-cyclotron waves and determine which propagation angles are most efficient in preferentially heating the minor ions within the considered low-frequency turbulent wave spectra. Our secondary goal is to examine the evolutionary path of particles in the configuration space set by the ion temperature anisotropies and their plasma $\beta$, see \citep{Bale:09,Maruca:12,Bourouaine:11,Bourouaine:13a, Wicks:13}. 
To address these problems, we perform 2.5D hybrid simulations with fluid electrons, kinetic protons and a minor component of differentially streaming $\alpha$ particles. We investigate the turbulent evolution, wave-scattering and dissipation of initial broad-band spectra of parallel and oblique Alfv\'en-cyclotron waves in collisionless finite plasma $\beta$ fast solar wind conditions. The simulation setup is motivated by the magnetic field power spectra and the plasma properties as \textit{in situ} observed by \textit{Wind} spacecraft near the Earth.
We start within the stable middle point where $T_{\perp i}/T_{\parallel i} =1$ and the plasma $\tilde \beta_i=0.33$ for both ion species, see definitions in the next section. This point is far from all linear plasma instabilities thresholds for both protons and the $\alpha$ particles \citep{Maruca:12, Bourouaine:13a} and yet the protons are cooled while the minor ions are heated in perpendicular direction and the parallel temperature of both species increases. Such cooling/heating can not be described by the Vlasov linear instability theory predictions. It is a nonlinear process, resulting from the turbulent wave-wave couplings and consequent wave-particle interactions with the minor ions, whose proper treatment requires direct numerical simulations.

The aim of this paper is to provide the first steps towards comparison and understanding of the efficiency of parallel versus oblique small-scale Alfv\'enic fluctuations for the preferential heating of minor ions in the fast solar wind. 
The initial spectrum resembles Alfv\'enic turbulence within the inertial range with a Kraichnan-type power spectral slope $\alpha=-3/2$. The spectral index for the fluctuations relates to in situ observations and has been calculated based on the Magnetic Field Instrument (MFI) magnetometer data on board of \textit{Wind} spacecraft in the undisturbed solar wind at 1AU. The electron temperature and the ion properties are obtained respectively from the the Three-Dimensional Plasma and Energetic Particle Investigation (3DP) instrument and the Solar Wind Experiment (SWE) particle detector on board of \textit{Wind}. The next section presents the particle data and the magnetic field measurements used in our model to construct a consistent initial state for the hybrid simulations. 
%
The description of the 2.5D hybrid model and the theoretical reconstruction of the initial wave spectrum are presented in Section~3 and Section~4. The results of the numerical simulations are revealed in Section~5. Discussion and concluding remarks are presented in Section~6.
%

\section{Initial state for the simulations -- \textit{Wind} data}
\label{Wind data}

In order to relate our model to real solar wind conditions we have constrained the initial state for the simulations to resemble in situ measurements in a finite-$\beta$ fast stream near the Earth. The initial plasma parameters for the simulations are based on analysis of \textit{Wind} spacecraft observations in the undisturbed fast solar wind at 1AU. In what follows below we will use the 92 seconds average of the \textit{Wind}/SWE (Faraday Cup) data to define the orientation of a mean magnetic field, as considered by \citep{Kasper:08,Kasper:13}. Thus we can define parallel and perpendicular components of all quantities with respect to the direction of the mean field. We will refer to the total proton plasma $\beta$ as the weighed sum of its parallel and perpendicular contributions, $\tilde{\beta_p} = 1/3(\tilde{\beta_{p \parallel}} + 2\tilde{\beta_{p \perp}})$, where the parallel proton plasma $\beta$ is defined in terms of the parallel component of the proton thermal velocity and the local Alfv\'en speed, $\tilde{\beta_{p \parallel}} \equiv v^2_\mathrm{th, p \parallel}/V^2_\mathrm{A}$. Analogously, the perpendicular part is given by $\tilde{\beta_{p \perp}} \equiv v^2_\mathrm{th, p \perp}/V^2_\mathrm{A}$. For isotropic plasmas the parallel and perpendicular components are equivalent: $\tilde{\beta_{p \parallel}} = \tilde{\beta_{p \perp}} = \tilde{\beta_p}$.
We use \textit{Wind}/MFI and SWE/Faraday Cup (FC) data to select a period with a given proton plasma $\beta$, temperature anisotropy and relative drift speed between the protons and the $\alpha$ particles. Once this interval is selected, we parametrize the magnetic field fluctuations, which are present there. The SWE/FC data samples are taken at 92 seconds, which sets the lowest temporal resolution from all instruments in our analysis. The SWE instrument \citep{Ogilvie:95} provides the proton density, temperature and temperature anisotropy, as well as the $\alpha$ particles density and the relative drift speed between the two ion species. The electron temperature is obtained from the 3DP instrument \citet{Lin:95} on board \textit{Wind} spacecraft. MFI data \citep{Lepping:95} provides vector magnetic field observations every 0.092 seconds. The different resolution of the instruments therefore leads to $10^3$ magnetic fields measurements within the 92 seconds period of one SWE/FC observation. The analyzed data is from 16$^{\mathrm{th}}$ of January 2008 at 15:19:09 UT and represents a long-lasting undisturbed fast wind stream during the recent solar minimum. The solar wind speed is 670 km/s. We have identified the time period with $\tilde {\beta_p} \approx 1$ and $T_{\parallel p}/T_{\perp p}\approx 1$ and selected the data point closest to (1,1) in the anisotropy-$\beta$ phase-space. Next, we select a scrolling window of 1000 surrounding points of MFI data and Fourier transform them to obtain spectral information about the magnetic fluctuations. The data analysis has been cross-checked with Morlet wavelet transformations and the same spectral properties are obtained. The trace of the power spectral tensor, the magnetic helicity and the variance anisotropy were computed using the local magnetic field direction, following the procedure explained in \citet{Wicks:12}. The normalized relative drift speed for the selected interval as obtained from analysis of the data from SWE/Faraday cup particle detector is $V_{\alpha p}/V_\mathrm{A} = 0.44$. The proton and the $\alpha$ particle number densities respectively are $n_p = 2.65 /cm^3$ and $n_{\alpha} = 0.09 /cm^3$. From the charge neutrality condition the electron number density amounts to $n_e = 2.83/ cm^3$. The initial temperature for the $\alpha$ particles has not been extracted from the data and for the rest of the modeling we will assume equal initial temperature for the two ion species. The amplitude of the mean magnetic field as obtained from the 92 seconds average of the \textit{Wind}/MFI measurements is $|\mathbf B_0| \approx$ 4 nT and the amplitude of the largest magnetic field fluctuation is about 0.2 nT. 
Figure~\ref{fig:mag_tr} shows the derived power spectral density of the magnetic field fluctuations in units of $nT^2/Hz$ as a function of the obtained wave frequency given in Hz (in the spacecraft frame) based on Fourier decomposition of the MFI data  (Morlet wavelet analysis of the data produces the same power spectral slopes.) The level of the instrument noise is over-plotted with a dashed line, and the spectra has been fitted by 2 separate power laws $f^{\alpha}$, where $\alpha$ represents the spectral slope. The noise level of the MFI instrument is not always constant nor visible in the spectra, but this particular estimate is in-line with the expectations of the Wind MFI instrument team (private communications with A. Koval). The original laboratory estimation for the instrument suggests a noise-level of $2\times 10^{-6} nT^2/Hz$. In reality the observed instrument noise floor is often slightly lower than that measured in the laboratory. For this study we have estimated a value of $8\times 10^{-7} nT^2/Hz$, which is roughly a factor of 3 lower than what is measured in the lab.
The initial lower frequency part of the spectrum, associated with large-scale fluctuations shows a $f^{-3/2}$ power law, typical for inertial range of a Kraichnan-type MHD turbulence, whereas the higher-frequency (smaller scales) part of the spectrum exhibits a steeper power slope $\alpha = -3$, characteristic for the dissipation range. Similar power slopes have been derived for a lower $\beta$ data surrounding the (1,0.33) point in anisotropy-$\beta$ phase-space (R. Wicks, private communications). In order to study direct cascade processes in the present modeling we will initialize the simulations with a fraction from the lower-frequency part of the magnetic power spectrum and let the kinetic micro-turbulence transfer the wave energy down to the ion scales, where the fluctuations can be absorbed due to wave-particle interactions.

\section{Hybrid code -- particles initialization}
\label{part_init}

The details of the 2.5D hybrid code and the numerical schemes used in our model have been presented in many previous studies, e.g. \cite{Ofman:07,OVM:11,OfmanVM:14,Maneva:15}. Within the present hybrid model the electrons are described as a charge-neutralizing massless isothermal fluid, and the ions are treated fully kinetically within the particle-in-cell approach. The code allows for two spatial coordinates, but computes all three components of the ion velocities, the current, and the electromagnetic fields. The electric and the magnetic fields are computed in Fourier space and periodic boundary conditions are assumed for both particles and fields solvers. The normalization of the temporal and spatial scales are introduced as follows: the simulation time is given in units of inverse proton gyro-frequency $\Omega_p^{-1}$ and the ion velocities are normalized to the Alfv\'en speed, $V_\mathrm{A} \equiv \frac{B_0}{\sqrt{\mu_0 n_em_p}},$ defined by the electron number density $n_e$, the proton mass $m_p$ and the magnitude of the homogeneous background magnetic field $B_0$. The length of the simulation box is given in units of the proton inertial length with equal resolution and number of grid points in both special dimensions, so that $L_x = L_y = 384 V_{\mathrm A}/\Omega_p.$ The simulations are performed with 256$\times$256 cells and 127 particles per cell per species; thus $> 16.6$ million particles are used. The time step is a small fraction of the proton gyro-period, $\Delta t = 0.02 \Omega^{-1}_p$. The size of each grid cell in $x$ and $y$ direction is the same and equals 1.5 proton inertial lengths. The code uses the expanding box model originally developed by \citep{Grappin:96}. Its application to hybrid models was first presented in \citep{Liewer:01} and has lately been widely used by many authors, e.g. \citet{Hellinger:05,Hellinger:06,OVM:11,Moya:12,Hellinger:13,Maneva:13a,OfmanVM:14,Maneva:15}. The detailed description of the expanding box model can be found in all of the given references and will not be repeated here. For the purpose of the present study we will only introduce the solar wind expansion parameter $\varepsilon = U_0 /R_0$, defined at $t_0 = 1/\Omega_p$ for solar wind outflows with constant velocity $U_0$ at distance $R_0$ from the Sun.
We load the code with homogeneous plasma density and isotropic drifting Maxwellian velocity distribution functions (VDFs) for both protons and the minor ions. The initial ion densities correspond to the \textit{Wind}/SWE measurements presented in the previous section, which also provide the initial proton temperature, as well as the relative drift speed used to initialize the simulations. The electron temperature is obtained from the \textit{Wind}/3DP instrument. Using the electron number density as a normalization, the normalized number density for the protons and the $alpha$ particles becomes: $n_p/n_e \approx 0.94;$ $n_{\alpha}/n_e \approx 0.03.$ The fluid electron plasma $\beta$ obtained from the 3DP observations is $\beta_e \equiv 2\mu_0n_e k_\mathrm B T_e/B_0^2 = 0.9$ The proton plasma $\beta$ is defined as the square of the normalized proton thermal speed and is also calculated from the \textit{Wind}/SWE data, see Section~\ref{Wind data}. We should note that the Alfv\'en speed used here is defined in terms of the total electron density $n_e$ and the proton mass $m_p$. In this sense the definition of the ion plasma $\beta$ for the ion species reads: $\tilde{\beta_i} = m_p/m_i 2\mu_0 n_e k_\mathrm B T_i/B_0^2$. The plasma $\beta$ for the minor ions is then assumed to be a fraction of the plasma $\beta$ for the protons, $\tilde{\beta_\alpha} \approx 0.08; \tilde{\beta_p} = 0.33$. This selection assures that the two ion species have the same temperature at the beginning of the simulations. The selected \textit{Wind} data interval shows equal parallel and perpendicular temperatures for the protons and the same is assumed for the $\alpha$ particles. The normalized initial relative drift speed between the two species is $V_{\alpha p} = 0.44 V_\mathrm{A}.$ The selected proton plasma-to-cyclotron frequency ratio for the simulations is set to $\omega_{\mathrm{p}i}/\Omega_{\mathrm{c}i} = 0.5\times10^4$, as calculated from the proton density and the magnetic field strength provided by the \text{Wind}/SWE and MFI instruments.

\section{Construction of the initial turbulent wave spectra}
\label{waves_init}

\begin{figure}
\centering
\includegraphics[width=0.7\textwidth]{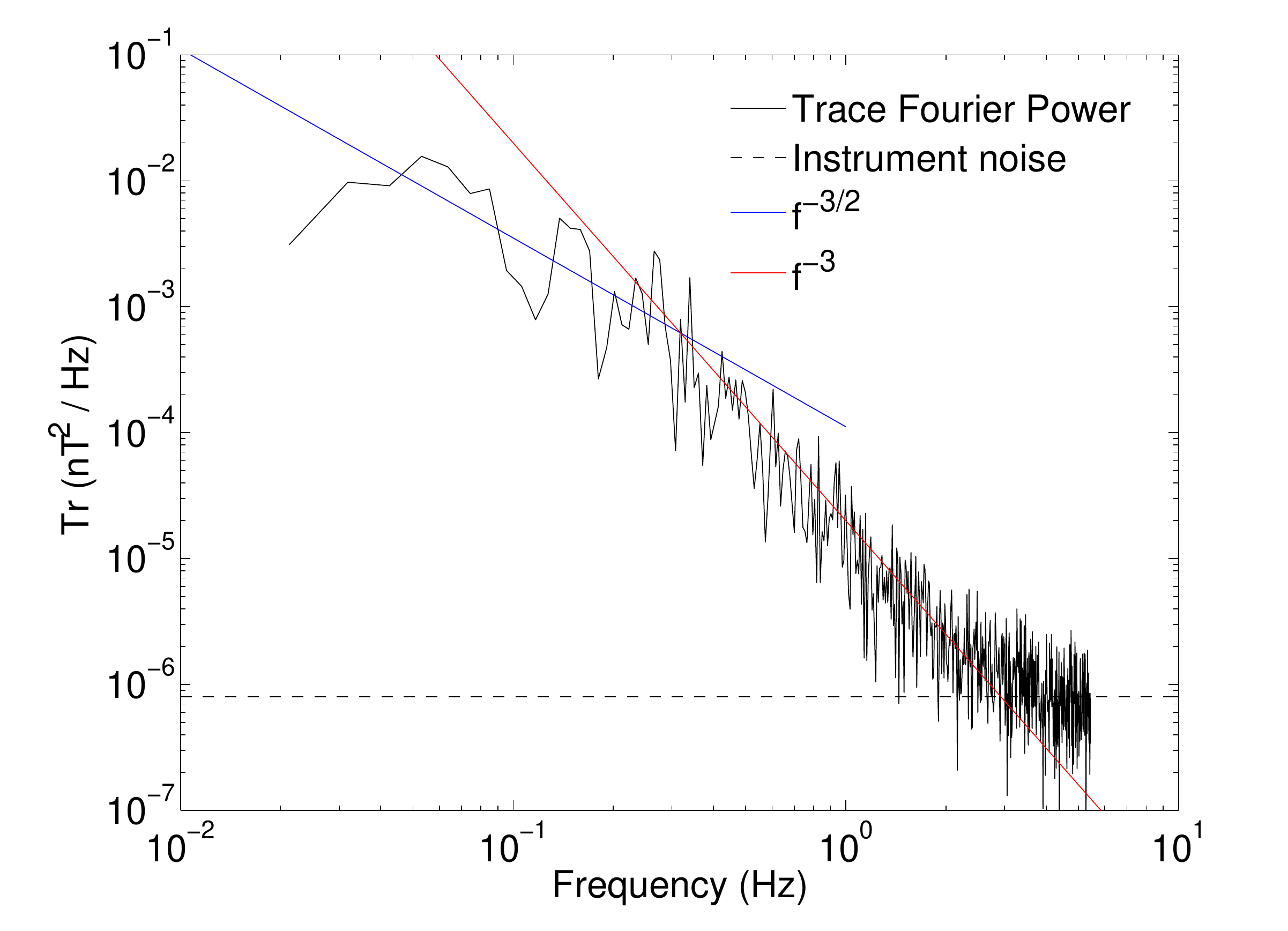}
\caption{Magnetic field power spectrum based on \textit{Wind}/MFI data fitted with two different power laws. The data is from 16th of January 2008, 15:19:09 UT.} 
\label{fig:mag_tr}
\end{figure}

In the previous section we presented the properties of the hybrid model used for this study and the initialization of the particle properties, such as thermal velocities, temperature anisotropies and relative drift speeds.
This section describes the theoretical reconstruction of an initial Alfv\'en-cyclotron turbulent wave spectra, which resembles the \textit{Wind}/MFI data presented in Figure~\ref{fig:mag_tr}. 
To construct the 2D oblique wave fronts needed for the 2.5D hybrid simulations we first generate a 1D broad-band wave spectra and rotate it in Fourier space to make it two-dimensional, preserving the energy and the magnetic field divergence free. The goal is to start loading the simulations with parallel Alfv\'en-cyclotron waves and vary the angle of propagation of the magnetic fluctuations and compare the evolution of the related anisotropic ion heating and differential acceleration for the different cases. \\
%
Let us assume that the homogeneous magnetic field background is in $x$ direction, $\mathbf{B}_0 = B_0\hat x$. The general reconstruction of a 1D broad-band spectrum of parallel-propagating magnetic field fluctuations follows the method developed in \citet{Vinas:14} and \citet{Vinas:84}
\begin{equation}
\boldsymbol{\mathbf{B}}\left(x\right)=\mathbf{B}_0 + Re\left[\sum^{k_{\rm{max}}}_{k_{\rm{min}}}\left(C_{k}^{L}\,\exp\left(i\varphi_{k}\right)
\left(\hat{\mathbf{y}}+i\mathbf{\hat{z}}\right)+C_{k}^{R}\,\exp\left(i\psi_{k}\right)\left(\mathbf{\hat{y}}-i\mathbf{\hat{z}}\right)\right)
\,\exp\left(ik_\parallel x\right)\,\right]\,\,\,.
\label{eq:EQ1}
\end{equation}
where ${k_{\rm{min}}}$, ${k_{\rm{max}}}$ are the lowest and highest wave-numbers characterizing the wave spectra and
\begin{equation}
\begin{array}{c}
\left|C_{k}^{L}\right|^{2}=\dfrac{\eta^{2}\, B_{0}^{2}}{2}
\left[\dfrac{k_\parallel^{-\alpha}\left(1+\sigma_{mk}\right)}{\sum^{k_{\rm{max}}}_{k_{\rm{min}}}k_\parallel^{-\alpha}}\right]\\
\\
\left|C_{k}^{R}\right|^{2}=\dfrac{\eta^{2}\, 
B_{0}^{2}}{2}\left[\dfrac{k_\parallel^{-\alpha}\left(1-\sigma_{mk}\right)}{\sum^{k_{\rm{max}}}_{k_{\rm{min}}}k_\parallel^{-\alpha}}\right]\,,
\end{array}\label{eq:EQ2}
\end{equation} 
are the Fourier amplitudes of the left-handed and right-handed modes. The 1D spectrum describes parallel wave propagation, $\mathbf{k}=k_{\parallel}\hat x,$ with random phases $\varphi_k$ and $\psi_k$ for each mode. In Eq.~\eqref{eq:EQ2} $\eta$ is the maximal spectral wave amplitude, $-1 \le \sigma_{mk}\le 1$ is the reduced magnetic helicity for each mode and $\alpha$ is the spectral index of a prescribed power-law power spectral profile. The spectrum conserves the wave energy, polarization, magnetic helicity and the cross-helicity in configuration space and Fourier space \citep{Vinas:14}. 
This general procedure of generating 1D magnetic field fluctuations was applied in a recent 2.5D simulation study \citep{Maneva:15} to construct 1D initial broad-band spectra of left-hand polarized parallel propagating Alfv\'en-cyclotron wave and follow their anisotropic nonlinear turbulent cascade. These simulations assumed that the entire simulation box is permeated by one-dimensional parallel propagating waves  and studied their role for heating and acceleration of minor ions in the solar wind. In order to initialize our 2.5D numerical simulations with 2D turbulent wave spectra in the present work at $y=0$ we start with the one-dimensional spectra calculated for all grid points $x_i$ along the direction of the background magnetic field. Next we transform the resulting 1D field $\mathbf{B}(x)$ into Fourier domain and perform a rotation of angle $\theta$ with respect to the $z$ axis to the obtain 2D oblique wave-vectors. The new wave vectors $\mathbf{k}'$ and the Fourier component of the complex magnetic field $\mathbf{B'_{k'}}$ acquired through the rotation are given by 
\begin{equation}
\mathbf{k}' \equiv k_\parallel' \hat x + k_\perp' \hat y  = k_\parallel \cos(\theta)\hat x + k_\parallel \sin(\theta)\hat y\,. 
\label{eq:EQ3}
\end{equation}
and
\begin{equation}
\mathbf{B'_{k'}} = -B_{k,y} \sin(\theta)\hat x + B_{k,y} \cos(\theta)\hat y + B_{k,z}\,,
\label{eq:EQ4}
\end{equation}
where
\begin{equation}
\begin{array}{c}
B_{k,y}=\frac{1}{2}\left(C_{k}^{L}\,\exp\left(i\varphi_{k}\right)+C_{k}^{R}\,\exp\left(i\psi_{k}\right)\right)\\
B_{k,z}=\frac{i}{2}\left(C_{k}^{L}\,\exp\left(i\varphi_{k}\right)-C_{k}^{R}\,\exp\left(i\psi_{k}\right)\right)\,.
\end{array}\label{eq:EQ5}
\end{equation}
We apply this procedure to transform the initial 1D parallel waves into 2D oblique fluctuations at a prescribed angle $\theta$, preserving the energy and the divergence-free property of the initial magnetic field $\nabla \cdot \mathbf{B}'=0$ (or $\mathbf{k}'\cdot\mathbf{B}' =0$ in Fourier domain) and obtain an oblique magnetic field Fourier spectrum with power in both parallel and perpendicular directions with respect to the background field.
In addition, using the new magnetic spectrum and wave-vectors we construct the bulk velocities spectra for each ion species, namely
\begin{equation}
\delta U_{s,k,y}\pm i U_{s,k,z}=-\left[\frac{\left(\omega/k-U_{\Vert s}\right)}{\left(1\mp\left(\omega-kU_{\Vert s}\right)/\Omega_{s}\right)}\right]\left(\frac{B'_{k,y}\pm i B'_{k,z}}{B_{0}}\right)\,,
\label{eq:EQ12}
\end{equation} 
where $s$ is the species index and the $+$(-) sign represent left(right) hand-polarized waves.

We repeat the process for each grid row until $y=y_{\rm{max}}$, generating $N_y$ oblique spectrum all with different random phases. 
To preserve the total energy initially considered we must normalize the Fourier coefficients to divide the energy of each perpendicular mode
to the number of repetitions made in the $k'_\perp$ direction (by construction the energy is well distributed within all the parallel modes $k'_\parallel$).
Finally, after the rotation of the entire 1D Fourier spectrum we obtain oblique two-dimensional spectra of magnetic field fluctuations. Simultaneously we rotate and compute the corresponding bulk velocities for each ion 
species as initially given by the parallel dispersion relation, used as our starting case. Applying an inverse 2D FFT transform we retrieve the desired oblique 2D waves spectrum of Alfv\'en-cyclotron waves with both magnetic and velocity field fluctuations, which we load in the two-dimensional spatial grid.

\begin{figure}
\centering
  \includegraphics[width=.49\linewidth]{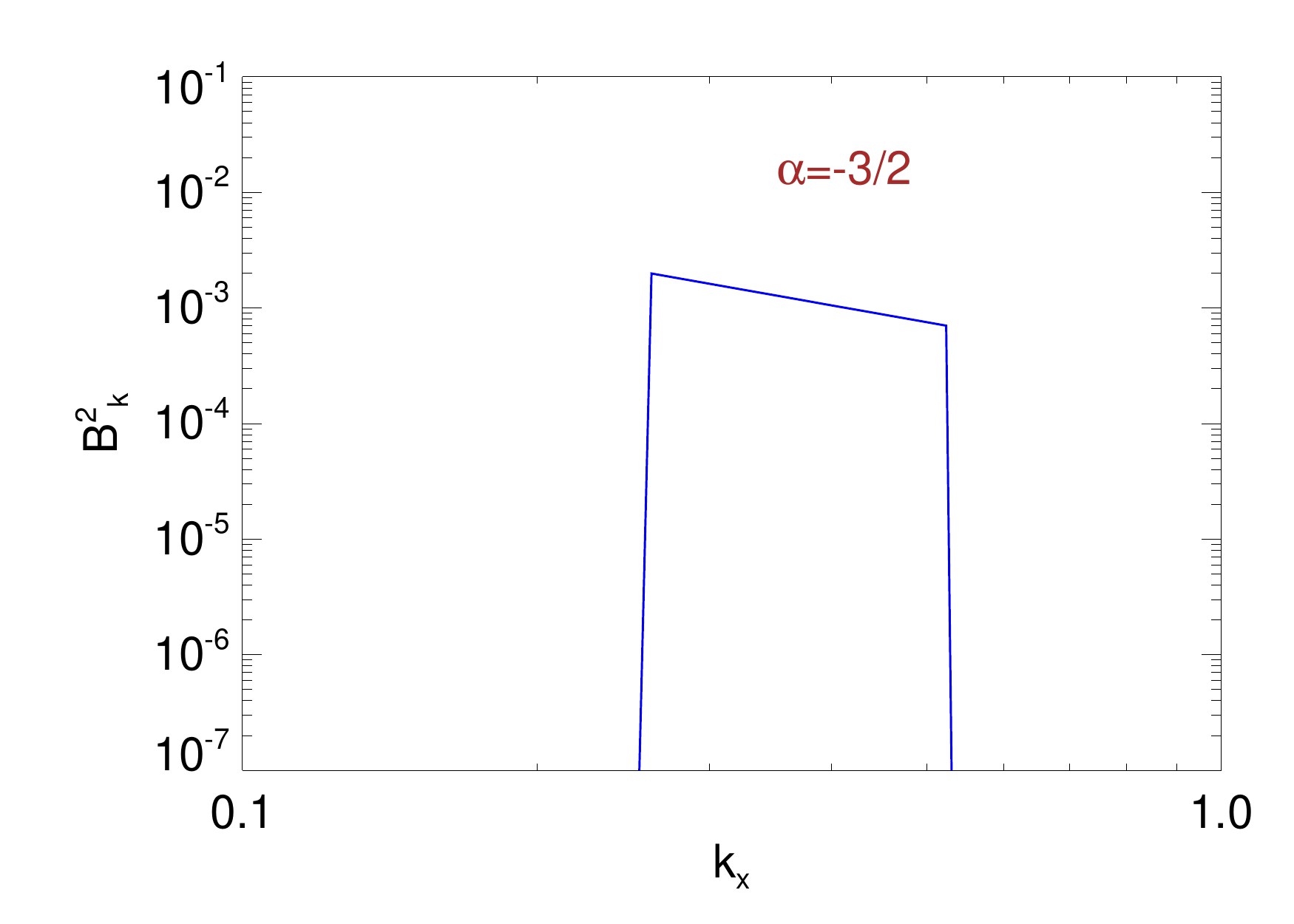}
  \includegraphics[width=.49\linewidth]{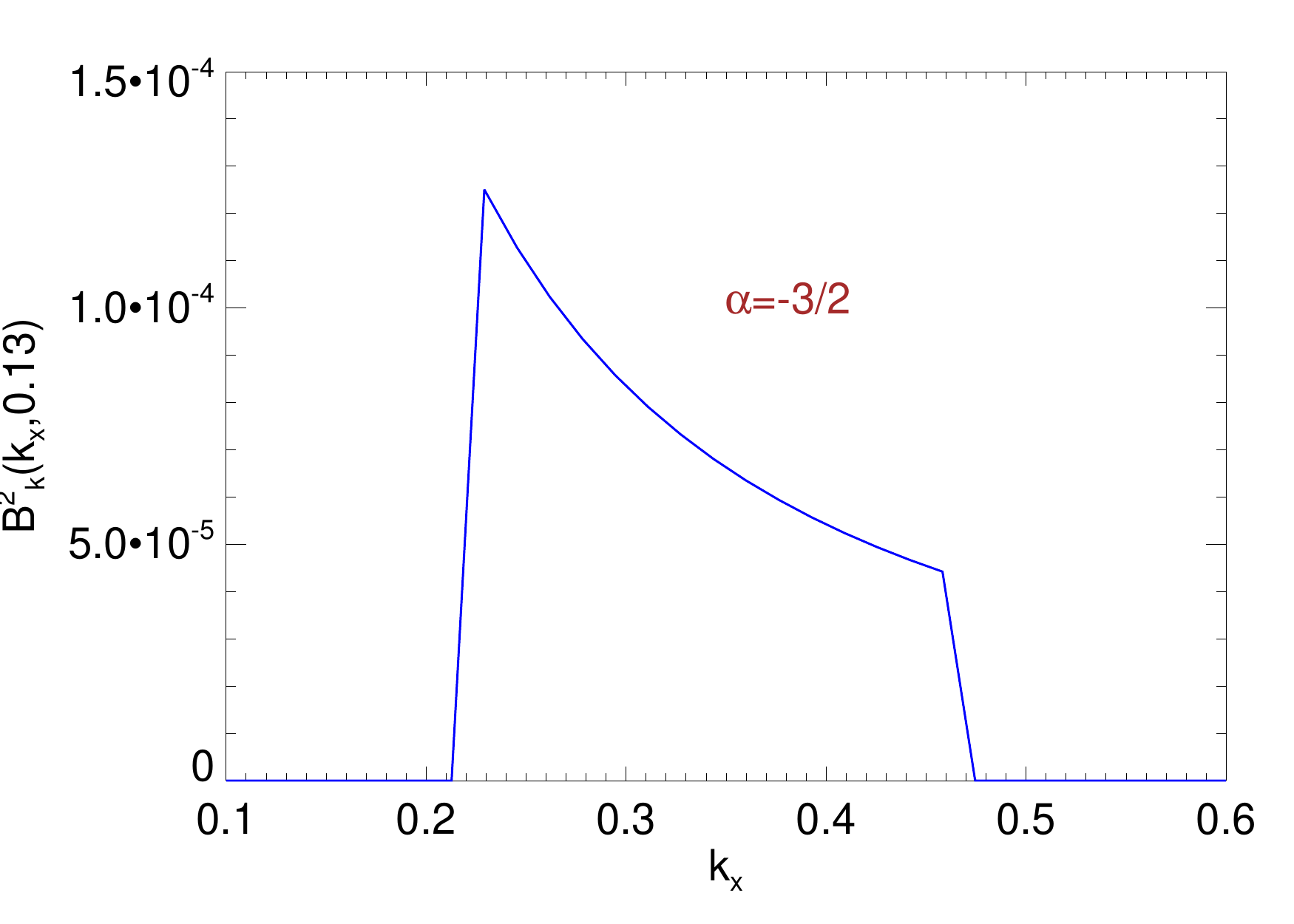}
\caption{Initial magnetic field power spectra as a function of the parallel wave-number for the case of parallel (left panel) and oblique (right panel) wave propagation at $\theta=30^\circ$. The oblique power spectra is calculated at $k_y=0.13 \Omega_p/V_\mathrm{A}$, but the spectral slope is the same for the entire interval of initial perpendicular wave-numbers $k_y \in [0.13,0.26] \Omega_p/V_\mathrm{A}$.}
\label{fig:bk2_0deg}
\end{figure}

\begin{figure}
\centering
\includegraphics[width=0.7\textwidth]{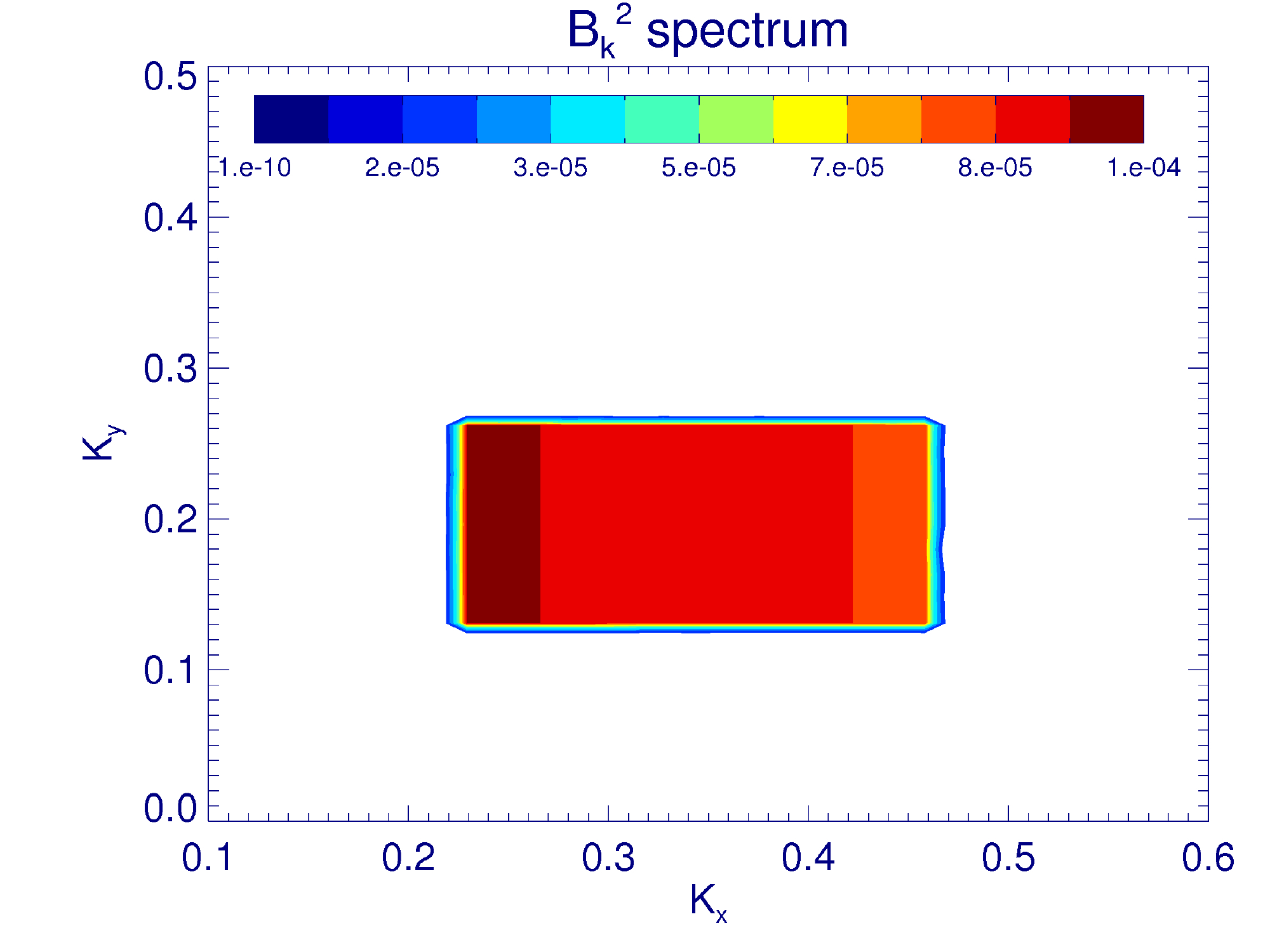}
\caption{Power spectral density for the initial two-dimensional magnetic field fluctuations as a function of the parallel, $k_x$, and perpendicular, $k_y$, wave-numbers for the case of oblique wave propagation at $\theta=30^\circ$. The magnetic field power is concentrated within the selected wave-numbers area presented in Table~1.} 
\label{fig:b_k_perp_par}
\end{figure}

\begin{figure}
\centering
\includegraphics[width=0.45\textwidth]{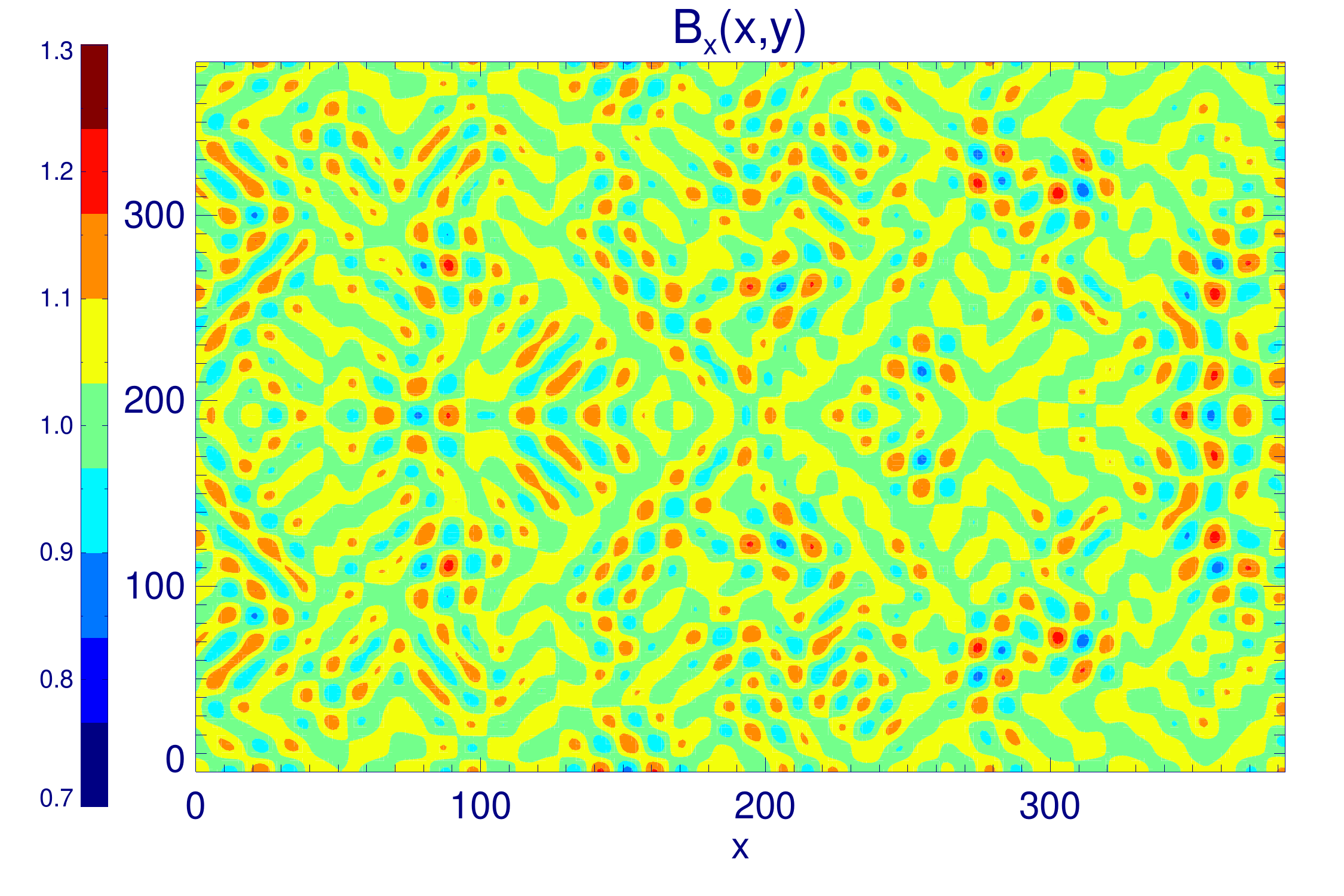}
\includegraphics[width=0.45\textwidth]{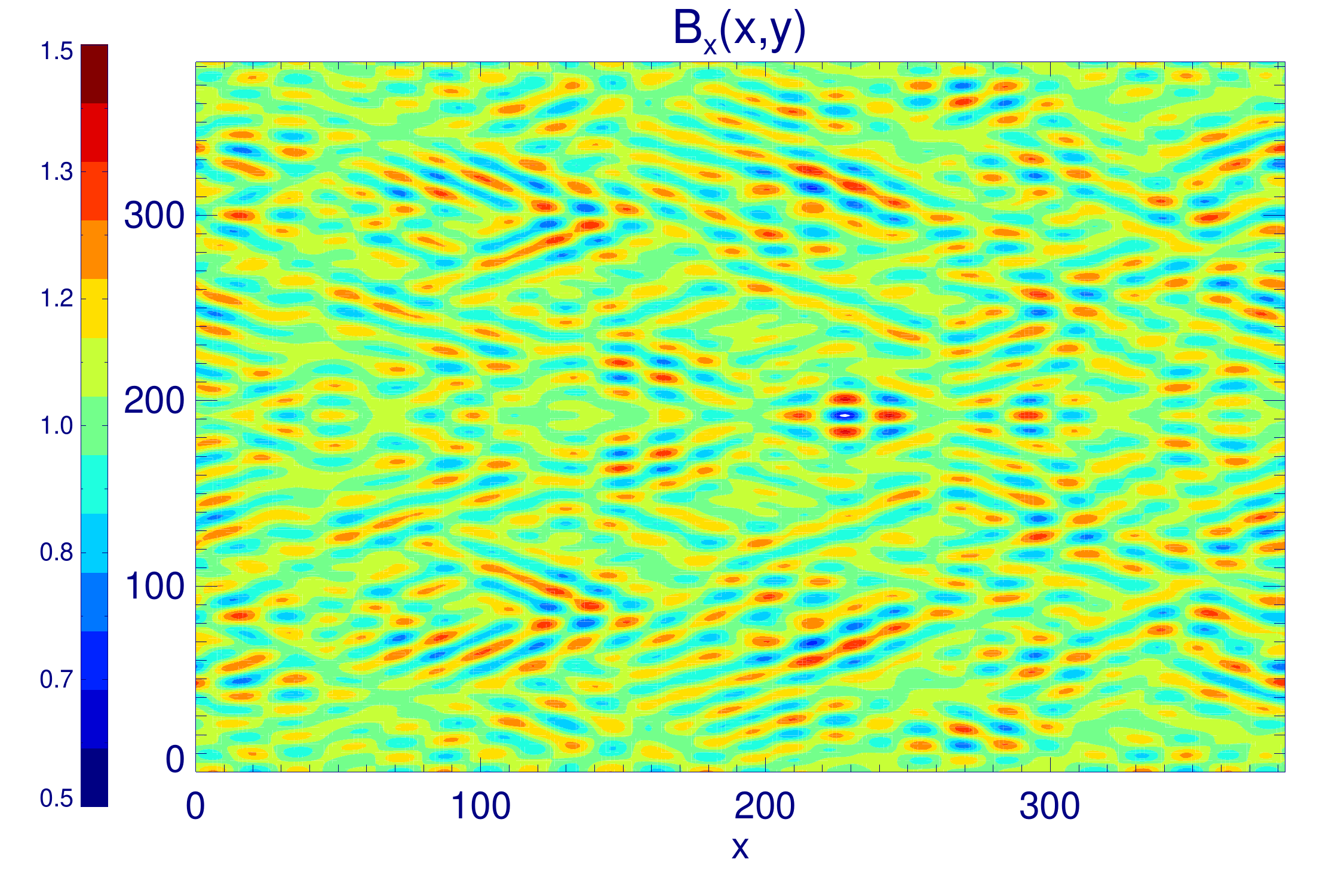}
\caption{Parallel component of the two-dimensional magnetic field in normalized units in real space. The field corresponds to oblique propagating waves at $\theta=30^\circ$ (left) and $\theta=60^\circ$ (right). The spatial coordinated are given in units of proton inertial length. The fluctuations are superimposed on the constant magnetic field background, whose amplitude is used for normalization. The fluctuations introduce inhomogeneity in the uniform background magnetic field.} 
\label{fig:bx_x_y}
\end{figure}

\begin{figure}
\centering
\includegraphics[width=0.47\textwidth]{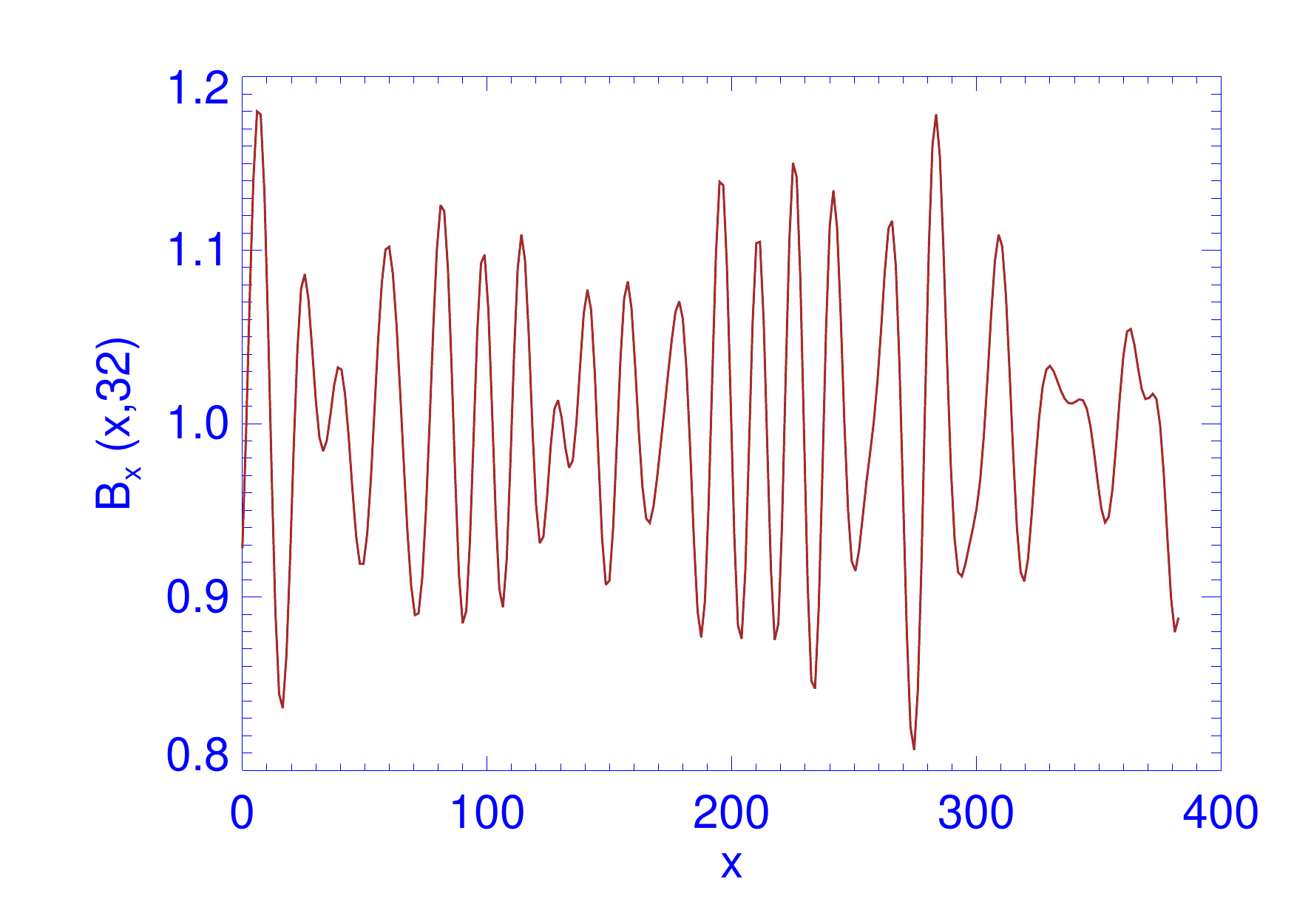}
\includegraphics[width=0.47\textwidth]{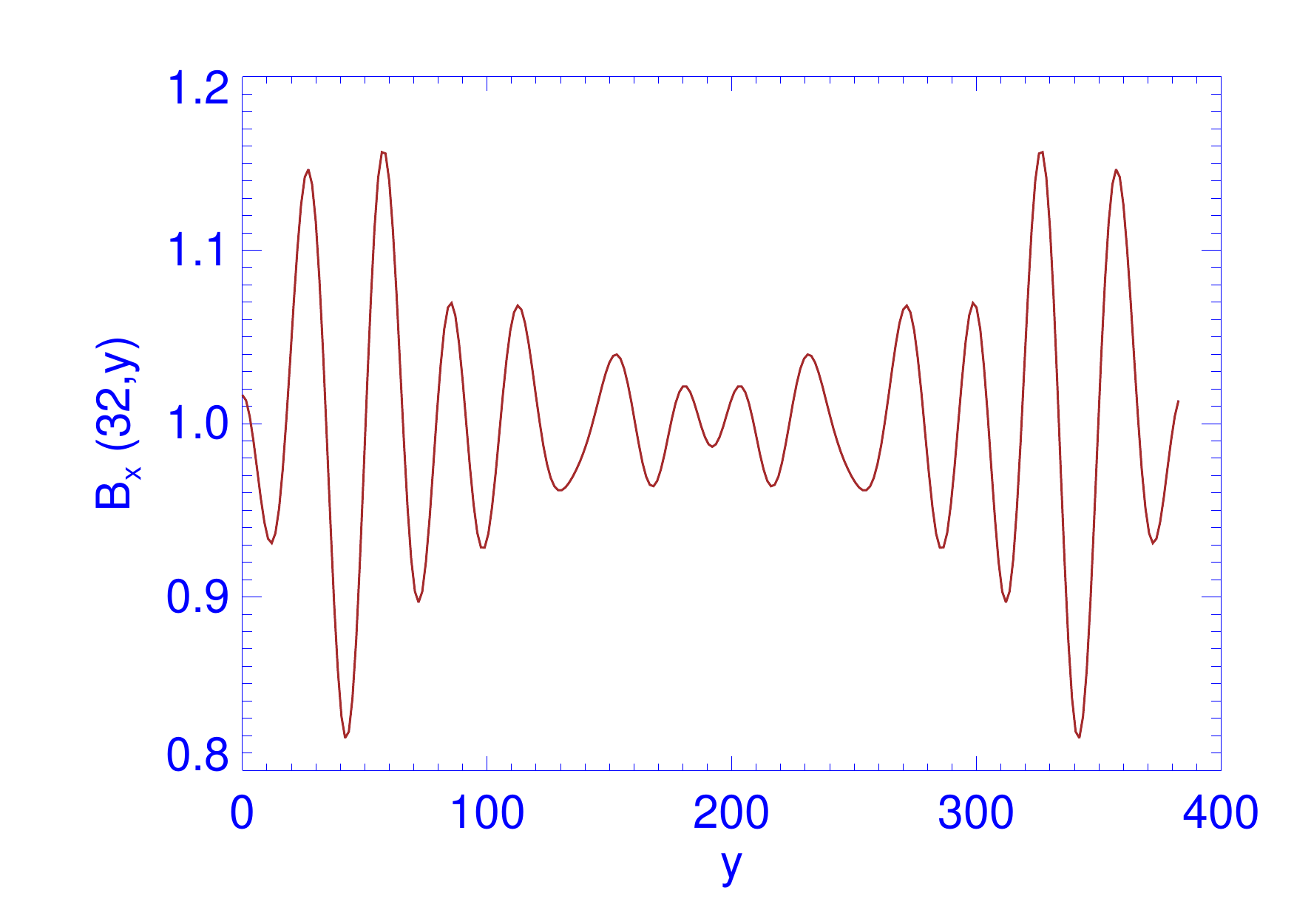}
\caption{Parallel component of the two-dimensional magnetic field in normalized units as a function of the parallel (left) and perpendicular (right) spatial coordinates, $x$ and $y$. The fluctuations correspond to obliquely propagating waves at $\theta=30^\circ$. The plots are complementary to the left panel of Fig.~\ref{fig:bx_x_y}.} 
\label{fig:bx_x}
\end{figure}

To compute the initial wave frequencies we solve the warm plasma dispersion relation for parallel wave propagation in a multi-species homogeneous plasma background. We select the left-hand circularly polarized waves and use them to construct the 1D velocity fields. The determinant of the dispersion tensor in the drifting multi-species magnetized plasma is given by the linear Vlasov theory \citep{Davidson:75,Sentman:81,Gary:02}
\begin{equation}
\mathrm{det} |\mathbf D| = k^2c^2 -\omega^2 - \sum_{s}\omega_{ps}\left(\frac{\omega - U_{s}k}{v_sk}Z(\xi^{+}_s) + (A_s-1)[1+\xi^{+}_{s}Z(\xi^{+}_{s})]\right). 
\end{equation}
The multi-species index $s$ in the summation stands for electrons, protons and $\mathrm{He}^{++}$ ions. The rest of the notations are as follows: $\omega_{ps}$, $v_s$, $U_s$ and $A_s$ are the plasma frequency, the spontaneous velocity, the bulk velocity and the temperature anisotropy for the $s$-th plasma species, $A_s = T_{\perp s}/T_{\parallel s}$. $Z(\xi)$ is the standard plasma dispersion function with Doppler-shifted frequency-dependend argument
\begin{equation}
\nonumber
Z(\xi^{+}_s) \equiv \pi^{-1/2} \int_{-\infty}^{\infty}\frac{\mathrm{e}^{-x^2}}{x-\xi^{+}_s}\mathrm dt, \quad \xi^{+}_{s} = \frac{\omega - U_sk + \Omega_s}{v_sk}.
\end{equation}
We look for solutions of the above equation for low-frequency waves, with frequencies below the lowest ion-cyclotron in the system, in this case the $\alpha$-cyclotron. For this study the electron inertia has been neglected and the electron contribution has been eliminated assuming charge neutrality and current conservation. The kinetic plasma dispersion relation is used in order to make sure that the selected initial wave spectrum is stable with respect to Vlasov theory plasma micro-instabilities. Once the frequencies for the parallel waves are obtained, the 1D fluctuations of the transverse velocity fields in the initially isotropic plasma are calculated from the multi-fluid dispersion relation: 
\begin{equation}
\delta \mathbf{V_{\perp i}}= - \frac{\omega - U_ik}{1-\left(\omega - U_ik\right)/\Omega_p} \delta \mathbf{B_\perp}/B_0
\end{equation}
Then the 1D velocity fluctuations for each species are rotated to obtain the 2D solution for the initial velocity fluctuations in configuration space.
The initial wave spectra considered here represent two-dimensional energy-preserving Alfv\'enic fluctuations and more precisely forward-propagating left-hand polarized Alfv\'en-cyclotron waves from the lower-frequency $\alpha$-cyclotron branch of the dispersion relation. The initial wave spectra demonstrated on Figure~\ref{fig:bk2_0deg} are chosen to resemble the end of the inertial range represented by the spectral slope $\alpha=-3/2$ from the \textit{Wind}/MFI spectrum, see Figure~\ref{fig:mag_tr}. The generated spectra consist of 17 modes and the total amplitude is $\delta B=0.2B_0$. For a fixed amplitude and spectral slope, we initialize the simulations with three different propagation angles, respectively $0\degr, 30\degr$ and $60\degr$, and study their influence on the minor ions. In the pure parallel case the waves have wave numbers $k_\parallel \in [0.26, 0.52] \Omega_p/V_\mathrm{A}$ and frequency range $\omega \in [0.22, 0.34] \Omega_p$. At $30\degr$ the spectrum has $k_\parallel \in [0.23, 0.46] \Omega_p/V_\mathrm{A}, \omega \in [0.21, 0.34] \Omega_p$ and $k_\perp \in [0.13, 0.26] \Omega_p/V_\mathrm{A}.$ The highly oblique case at $60\degr$ has $k_\parallel \in [0.13, 0.26] \Omega_p/V_\mathrm{A}, \omega \in [0.13, 0.26]\Omega_p$ and $k_\perp \in [0.23, 0.46] \Omega_p/V_\mathrm{A}.$ 
Figure~\ref{fig:bk2_0deg} describes the reconstructed initial magnetic field power spectra as a function of the parallel component of the wave-vector. The left panel shows the case of strictly parallel propagating waves, where $\mathbf k=k_x \hat x$. The right panel shows the $k_x$-dependence of the magnetic spectra at fixed perpendicular component $k_y=0.13\Omega_p/V_\mathrm A$ for the case of oblique wave propagating at $\theta = 30^\circ$. The same spectral slope $\alpha=-3/2$ was used in both cases. 
Since the initial angle is not too large though, most of the power still remains in the parallel direction and the power law retains its slope. This is no longer true when highly oblique fluctuations at $\theta=60^\circ$ are considered. Yet, the magnetic field spectra as a function of the full wave-vector remains the same for all angles of propagation and recovers the parallel case shown on the left panel of the figure.
We should note that the initial power spectral law for the magnetic field fluctuation is constructed in $k$-space, whereas the observational data provides information about the power spectra in frequency domain. For the low-frequency MHD part of the Alfv\'enic spectrum the dispersion relation is linear (waves propagate at the Alfv\'en speed) and the two descriptions become identical. For the intermediate scales at the beginning of the dispersive part of the spectrum, which we address in our study, the spectral slope of the magnetic field fluctuations in frequency domain which results from $k_\parallel^{-3/2}$ reconstruction will be flatter with a larger corresponding power law exponent $\alpha > -1.$ 
Figure~\ref{fig:b_k_perp_par} depicts the 2D power spectral density of the initial magnetic field fluctuations for the case of oblique Alfv\'en-cyclotron waves propagating at $\theta=30^\circ$. The figure shows the squared magnitude of the magnetic field fluctuations in Fourier space as a function of the parallel $k_x$ and perpendicular $k_y$ components of the wave-vector. The initial wave power is restricted in the wave-number space described above and presented in Table~1. The following two figures illustrate the corresponding initial magnetic field in configuration space for the case of oblique wave propagation at $\theta=30^\circ$ and $\theta=60^\circ$. Figure~\ref{fig:bx_x_y} shows the parallel component of the magnetic field fluctuations as a function of the parallel $\hat x$ and perpendicular $\hat y$ spatial coordinates, forming the simulation box. The fluctuations are normalized with respect to the magnitude of the background field $B_0$ and the spatial coordinates are given in terms of proton inertial length $d_i\equiv c/\omega_{pl} = V_\mathrm{A}/\Omega_p,$ where $c$ is the speed of light in vacuum and $\omega_{pl}$ is the proton plasma frequency. The magnetic field in configuration space for the case of $\theta=30^\circ$ is given by the inverse Fourier transform of the initial fluctuations, presented in Figure~\ref{fig:b_k_perp_par}. The background field has not been subtracted, which shifts the magnitude of the fluctuations, forcing them to oscillate around the mean normalized value of 1. We should note that in the case of strictly parallel wave propagation there are no fluctuation along the background magnetic field. The non-zero fluctuations in $B_x$ are associated with the presence of oblique waves. 
Figure~\ref{fig:bx_x} is a further illustration of the parallel magnetic field oscillations in the case of oblique wave propagation at $\theta=30^\circ$ as an example. The two panel represent a cut of the parallel magnetic field component at $y=32 d_i$ (left panel) and $x=32d_i$ (right panel). As these positions the initial waves are still clearly present in both parallel direction along $x$, and perpendicular direction along $y$ regardless the limited range of selected wave-numbers for the reconstruction of the initial broad-band spectra, see Figures~\ref{fig:b_k_perp_par} and \ref{fig:bx_x_y}.\\
In the next section, we will present the influence of the initial wave spectra for the anisotropic ion heating for the different propagation angles discussed above. We will discuss the wave-scattering and the evolution of the ion velocity distribution functions for protons and $\alpha$ particles in the electromagnetic field of the waves. We will show the evolution of the initial magnetic field and plot the end-stage electromagnetic dispersion relation to analyze the scattering of the initial spectra along with the generation of new wave modes.

\section{Results}

\begin{figure}
\centering
\includegraphics[width=0.87\textwidth]{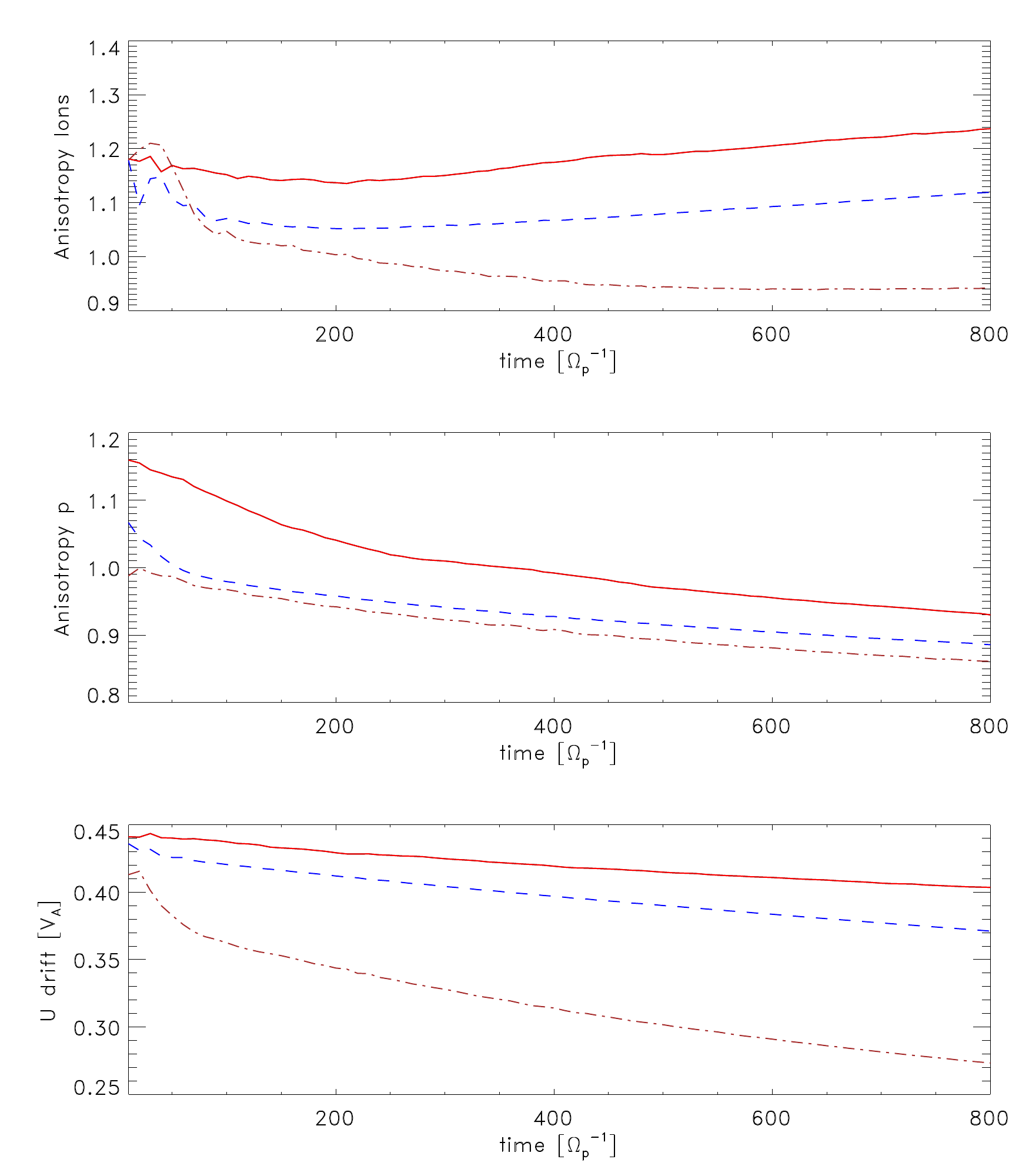}
\caption{Temporal evolution of the ion temperature anisotropies $T_{\perp i}/T_{\parallel i}$ and relative drift speed for the case of parallel (solid red line) and oblique wave propagation at $\theta=30^\circ$ (blue dashed line) and $\theta=60^\circ$ (brown dash-dotted line).} 
\label{fig:anisodr}
\end{figure}

\begin{figure}
\centering
\hspace{-0.25mm} \includegraphics[height=0.3\textheight]{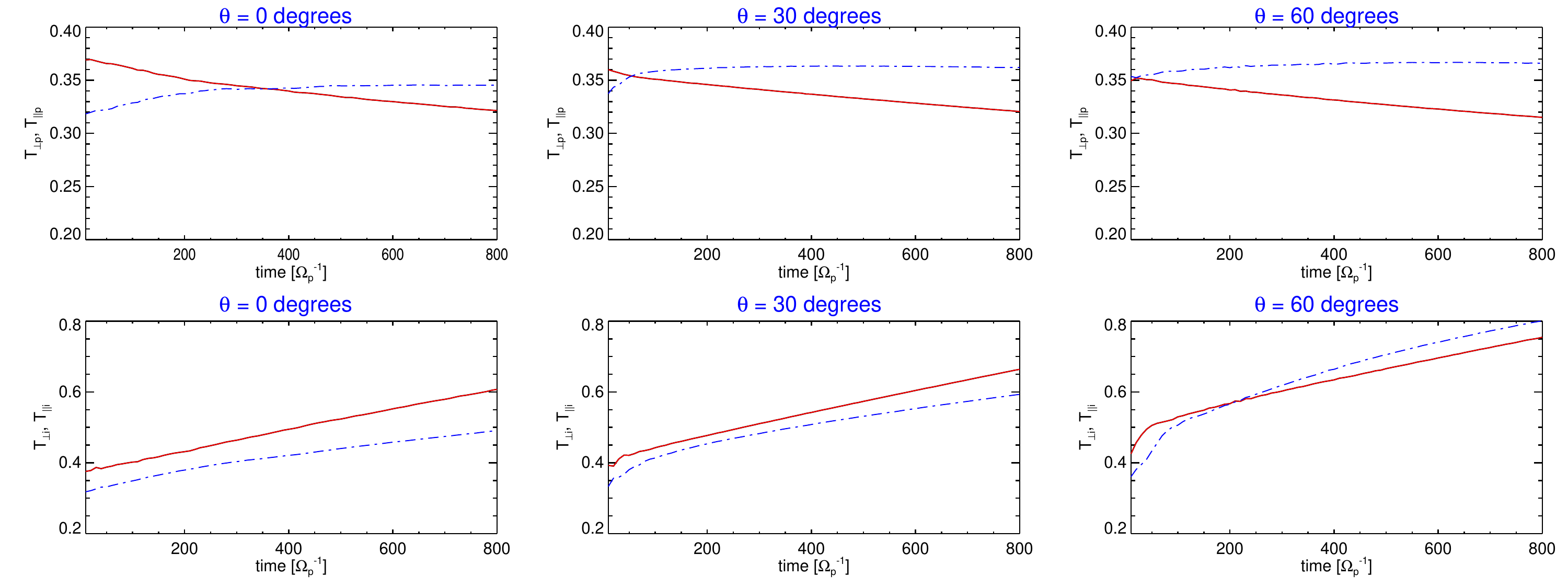}
\caption{Evolution of the parallel (dash-dotted blue lines) and perpendicular (solid red) ion temperature components for $\alpha$ particles (top row) and protons (lower row). The different panels show the different perpendicular heating for the minor ions induced by the 3 different initial wave spectra of parallel (first column) and oblique waves at $\theta=30^\circ$ (second column) and at $\theta=60^\circ$ (third column).} 
\label{fig:temp}
\end{figure}

\begin{figure}
\centering
\begin{minipage}{.32\textwidth}
  \centering
  \includegraphics[width=.95\textwidth]{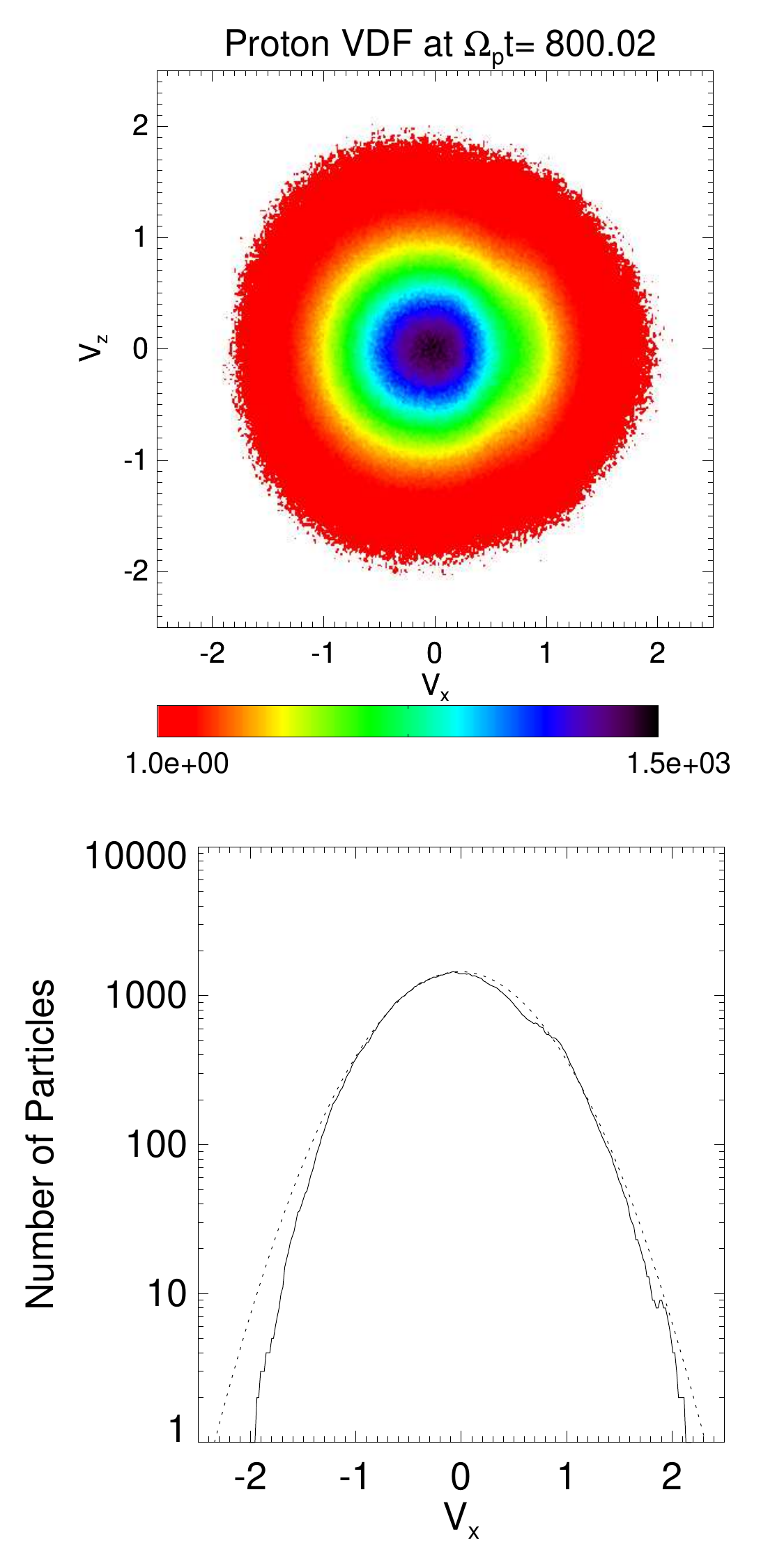}
\end{minipage}
\begin{minipage}{.32\textwidth}
  \centering
  \includegraphics[width=.95\textwidth]{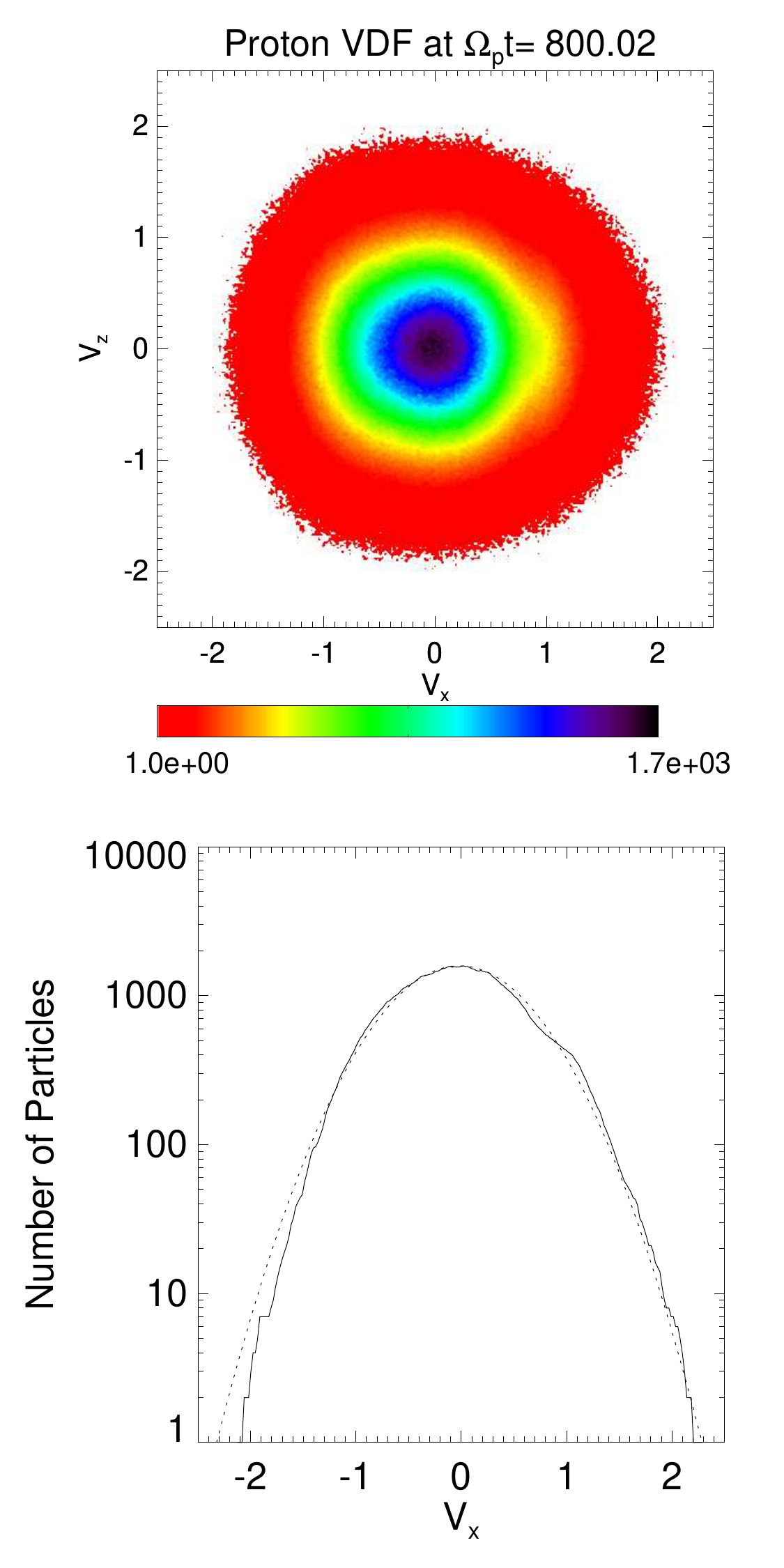}
\end{minipage}
\begin{minipage}{.32\textwidth}
  \centering
  \includegraphics[width=.95\textwidth]{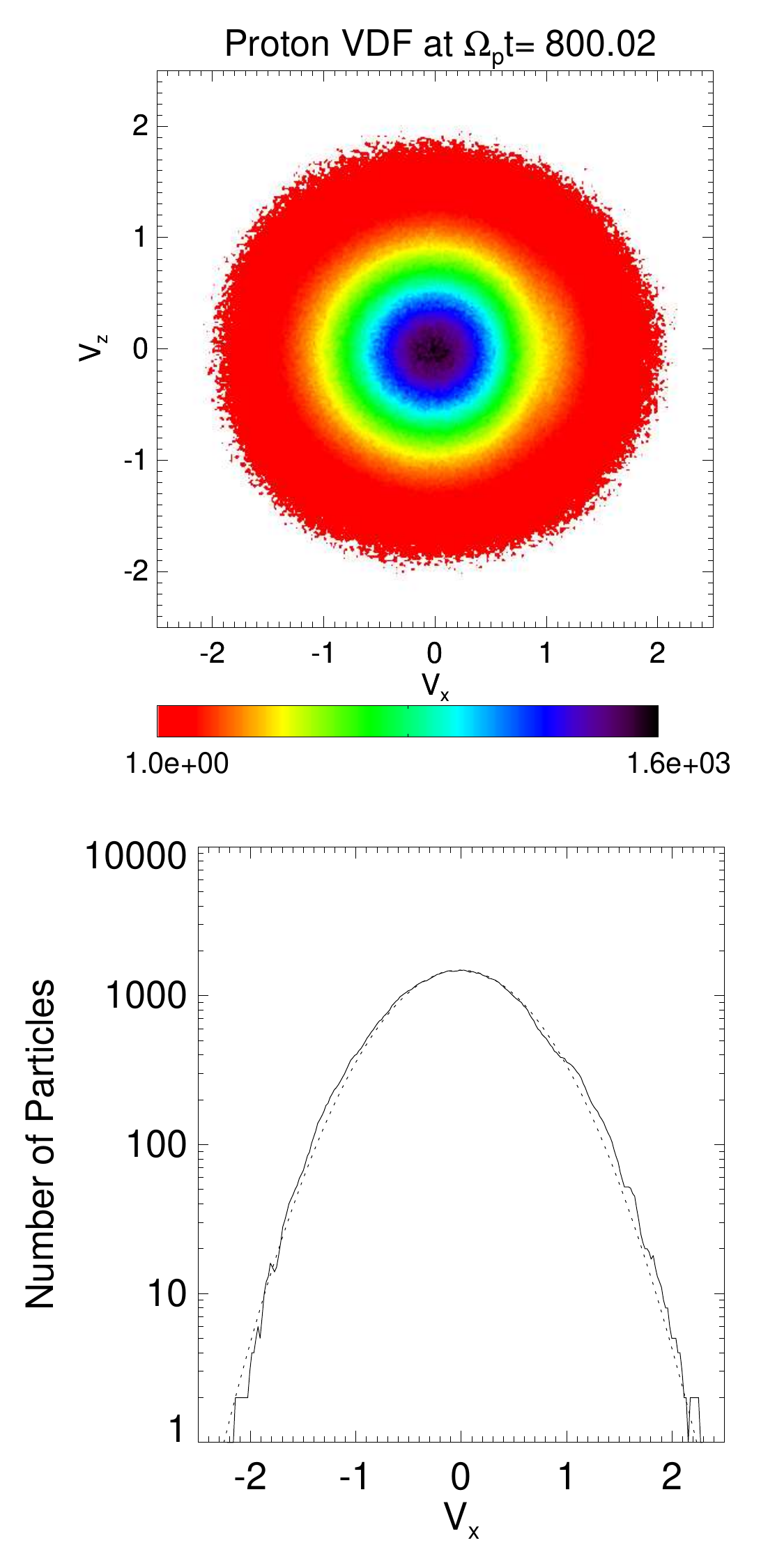}
\end{minipage}
\caption{Snapshots with the velocity distribution functions for protons for initially parallel waves (left panel), and oblique waves at $\theta=30^\circ$ (middle panel) and at $\theta=60^\circ$ (right panel).}
\label{fig:vdf_p0_30_60}
\end{figure}

\begin{figure}
\centering
\begin{minipage}{.32\textwidth}
  \centering
  \includegraphics[width=.95\textwidth]{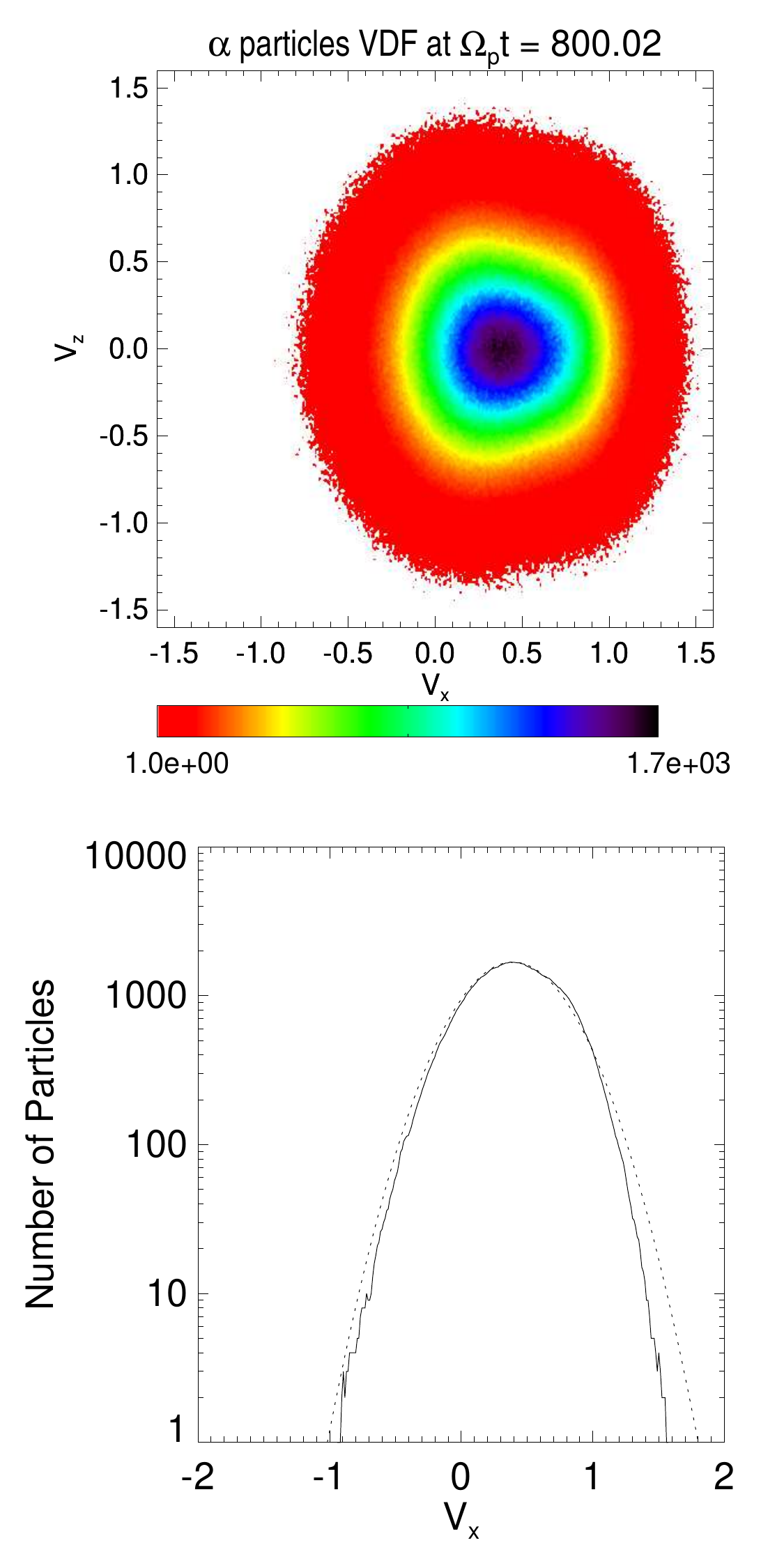}
\end{minipage}
\begin{minipage}{.32\textwidth}
  \centering
  \includegraphics[width=.95\textwidth]{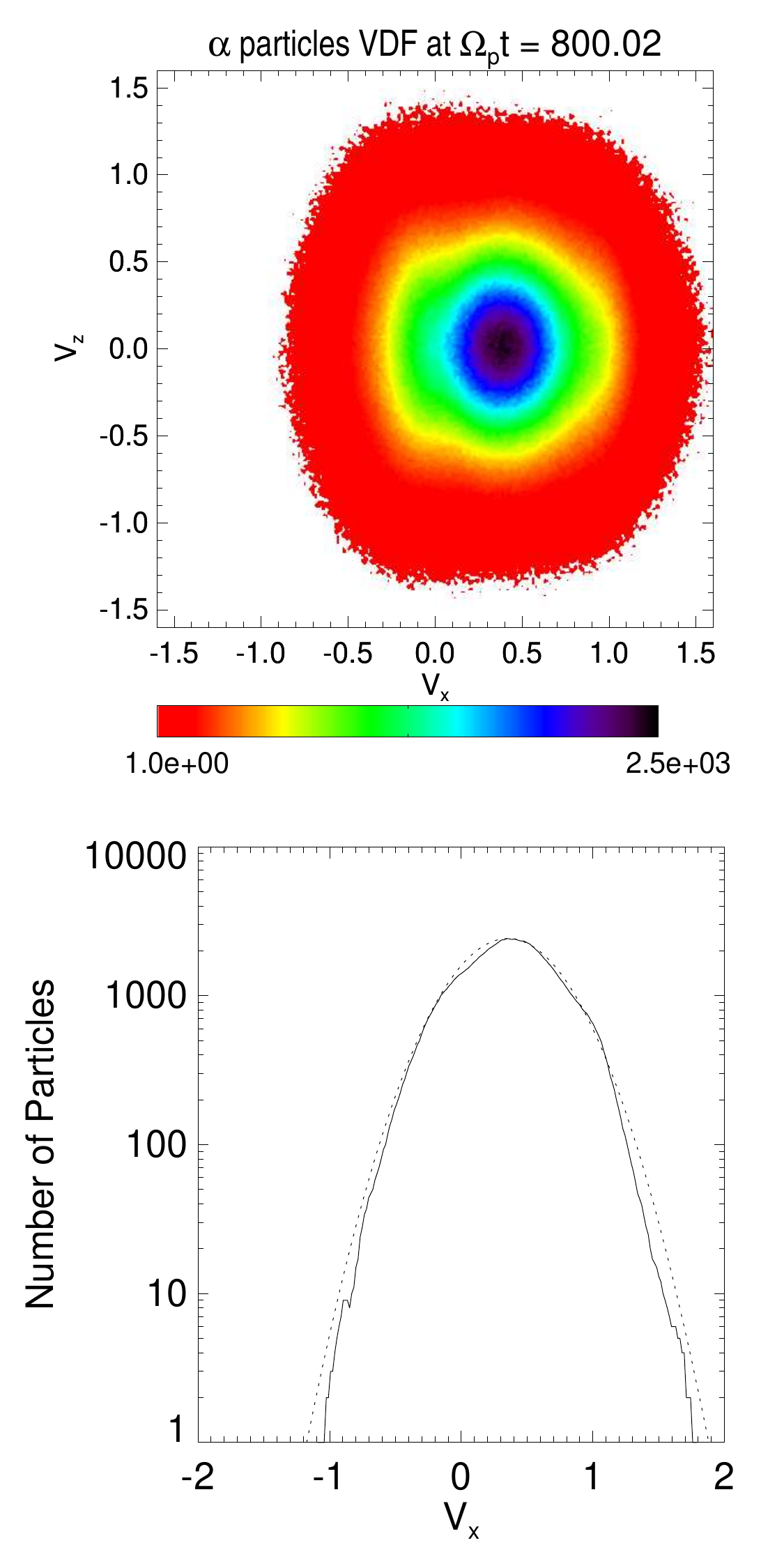}
\end{minipage}
\begin{minipage}{.32\textwidth}
  \centering
  \includegraphics[width=.95\textwidth]{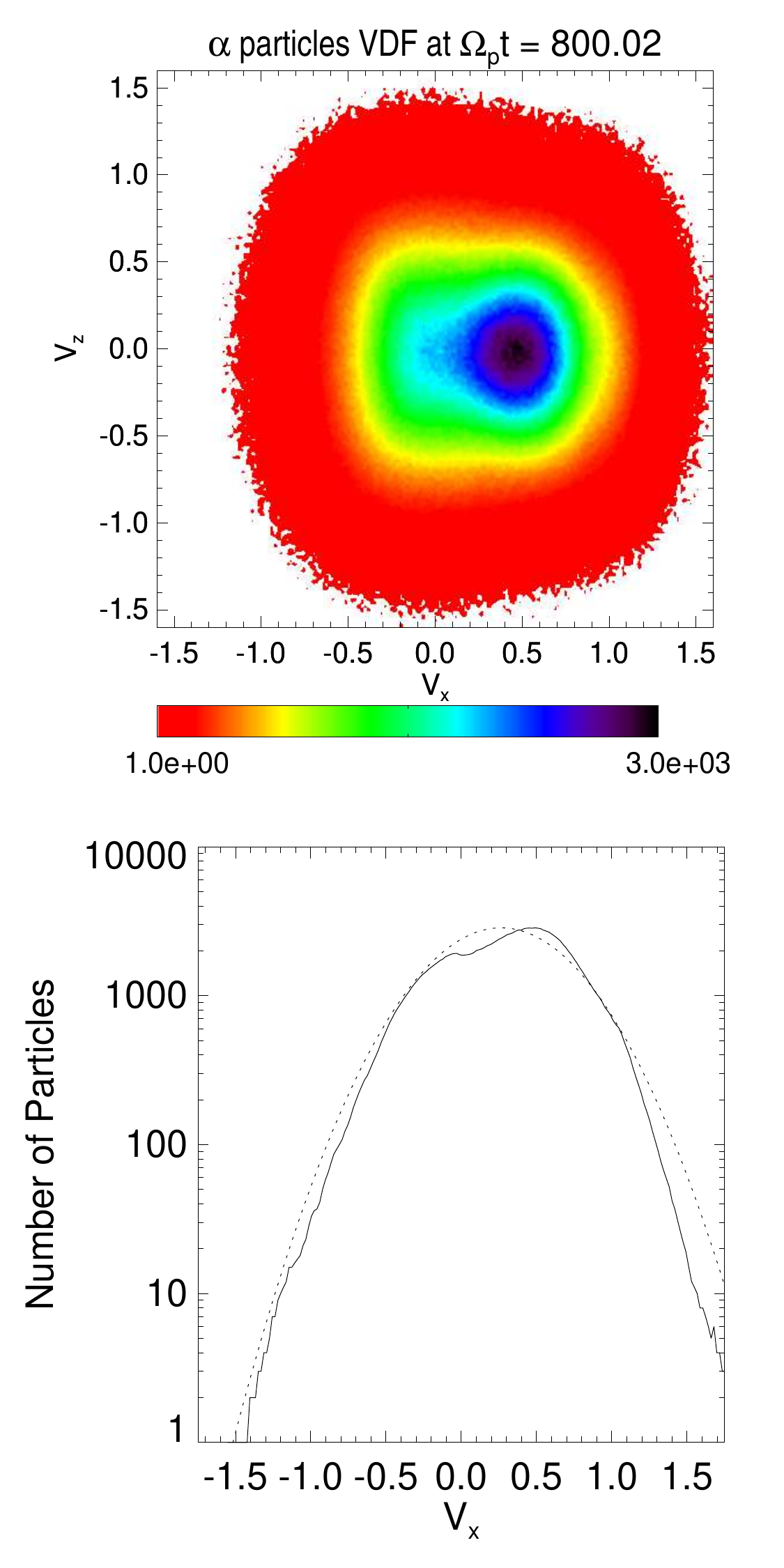}
\end{minipage}
\caption{Snapshots with the velocity distribution functions for $\alpha$ particles for initially parallel (left panel) and oblique waves at $\theta=30^\circ$ (middle panel) and at $\theta=60^\circ$ (right panel).}
\label{fig:vdf_a0_30_60}
\end{figure}

\begin{figure}
\centering
\includegraphics[width=0.6\textwidth]{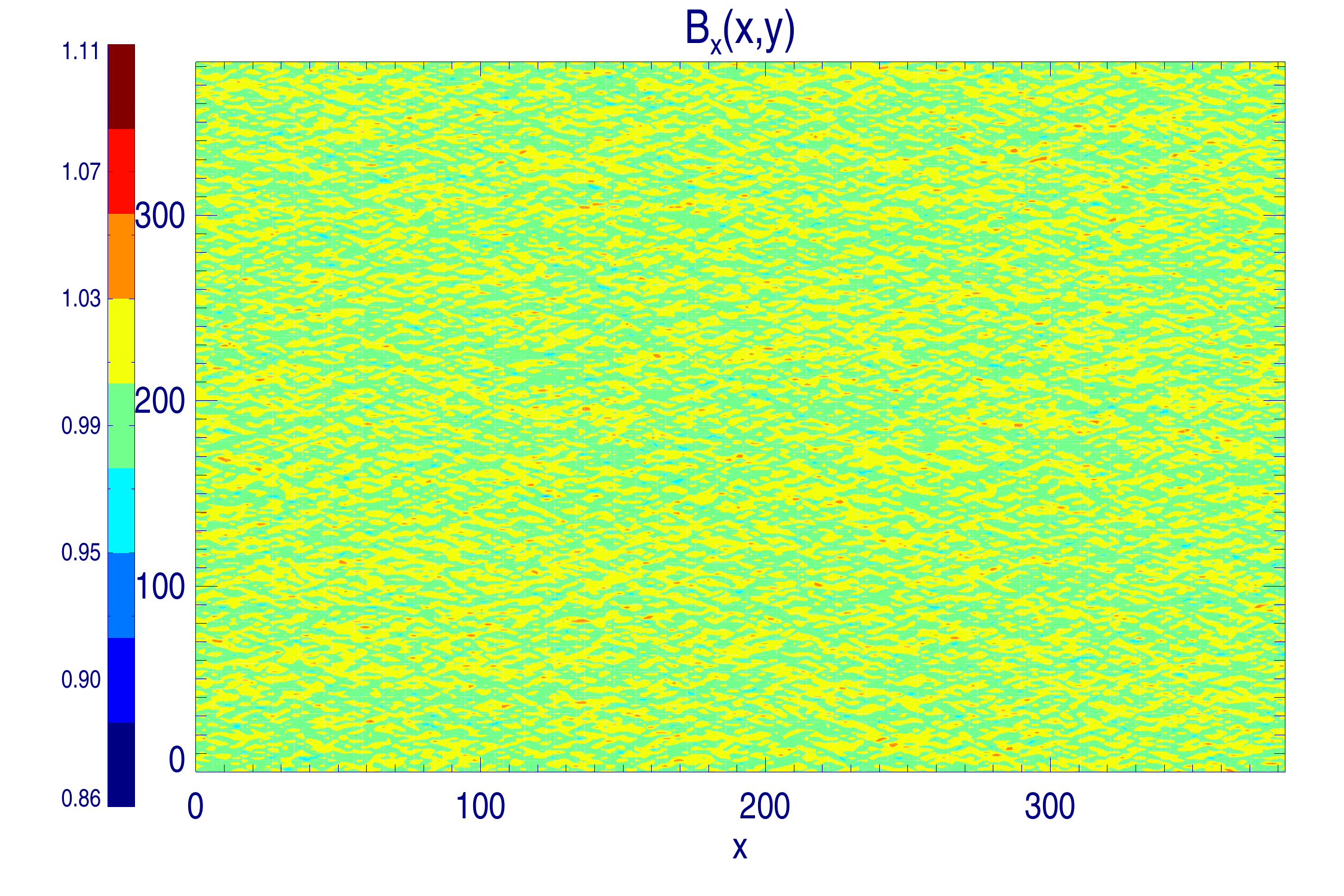}
\includegraphics[width=0.6\textwidth]{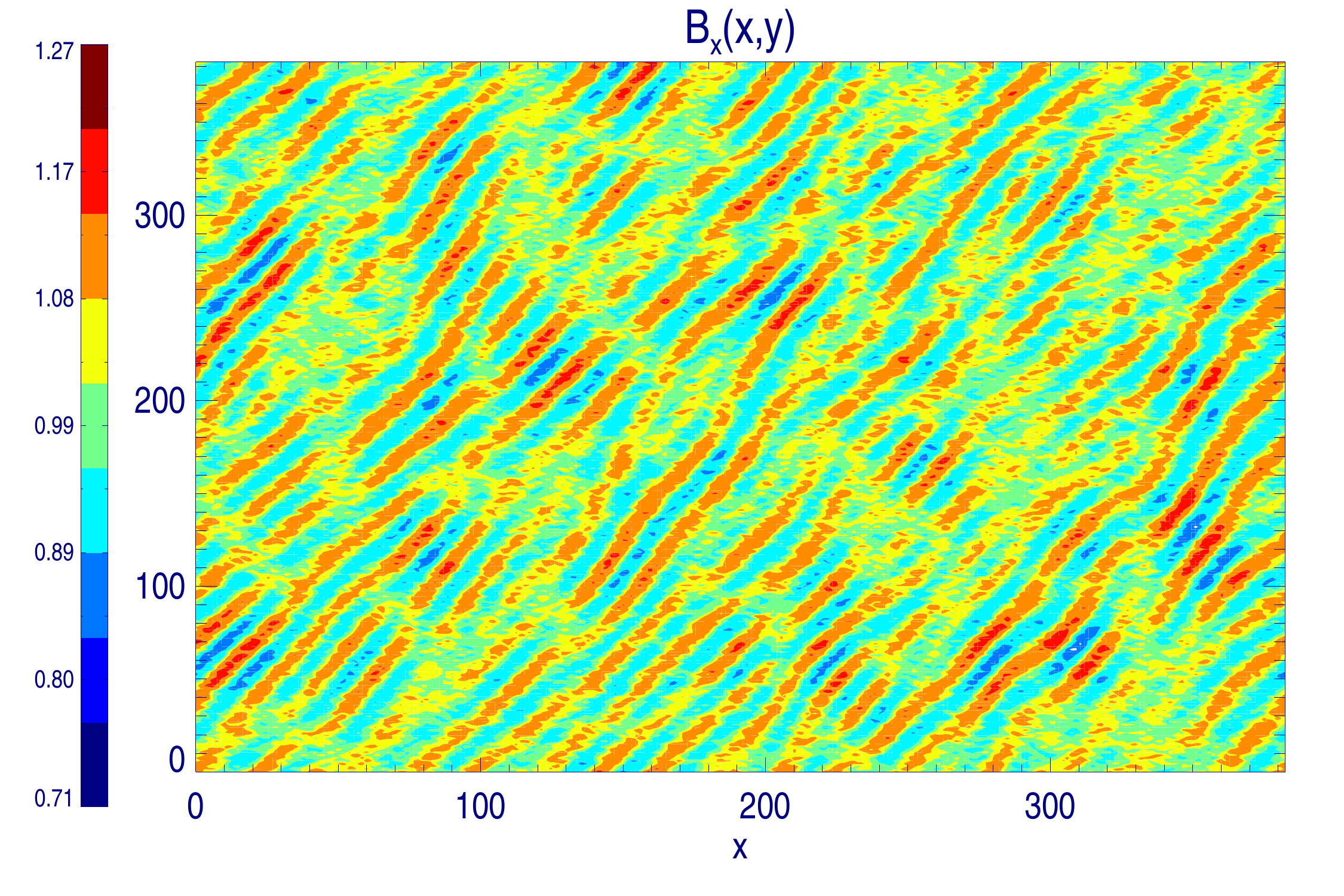}
\includegraphics[width=0.6\textwidth]{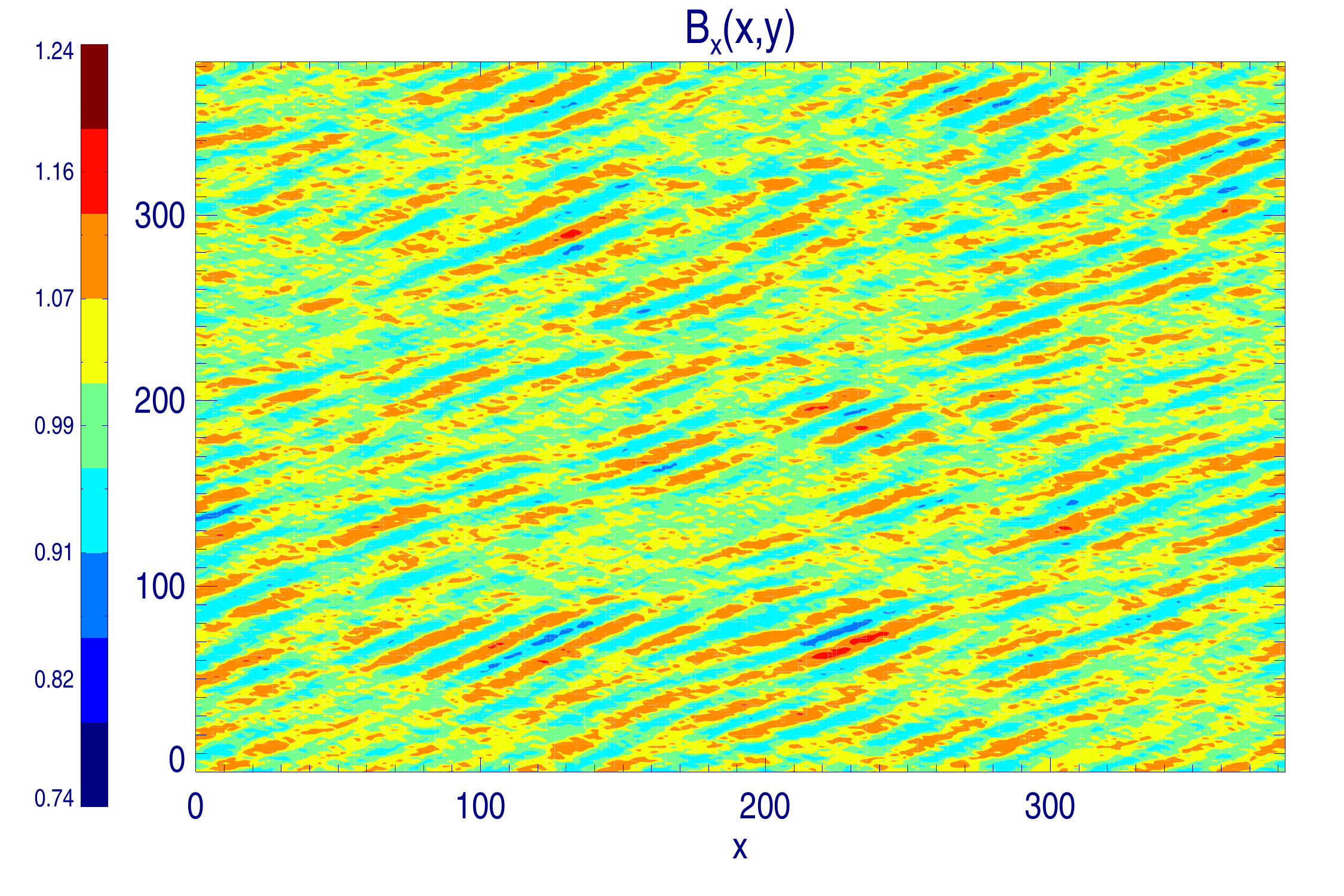}
\caption{2D magnetic field fluctuations along the background field at $\Omega_pt=800$ as a function of normalized coordinates in configuration space. The top panel relates to oblique fluctuation formed from the initially parallel wave spectra. The middle panel shows the end-state magnetic field variations for the case of oblique waves propagating at $\theta=30\deg$. The lower panel represents the reconfiguration of the magnetic field fluctuations for the case of initial oblique wave propagation at $\theta=60\deg$.} 
\label{fig:bx_x_y_end_sim}
\end{figure}

\begin{figure}
\centering
\includegraphics[width=0.7\textwidth]{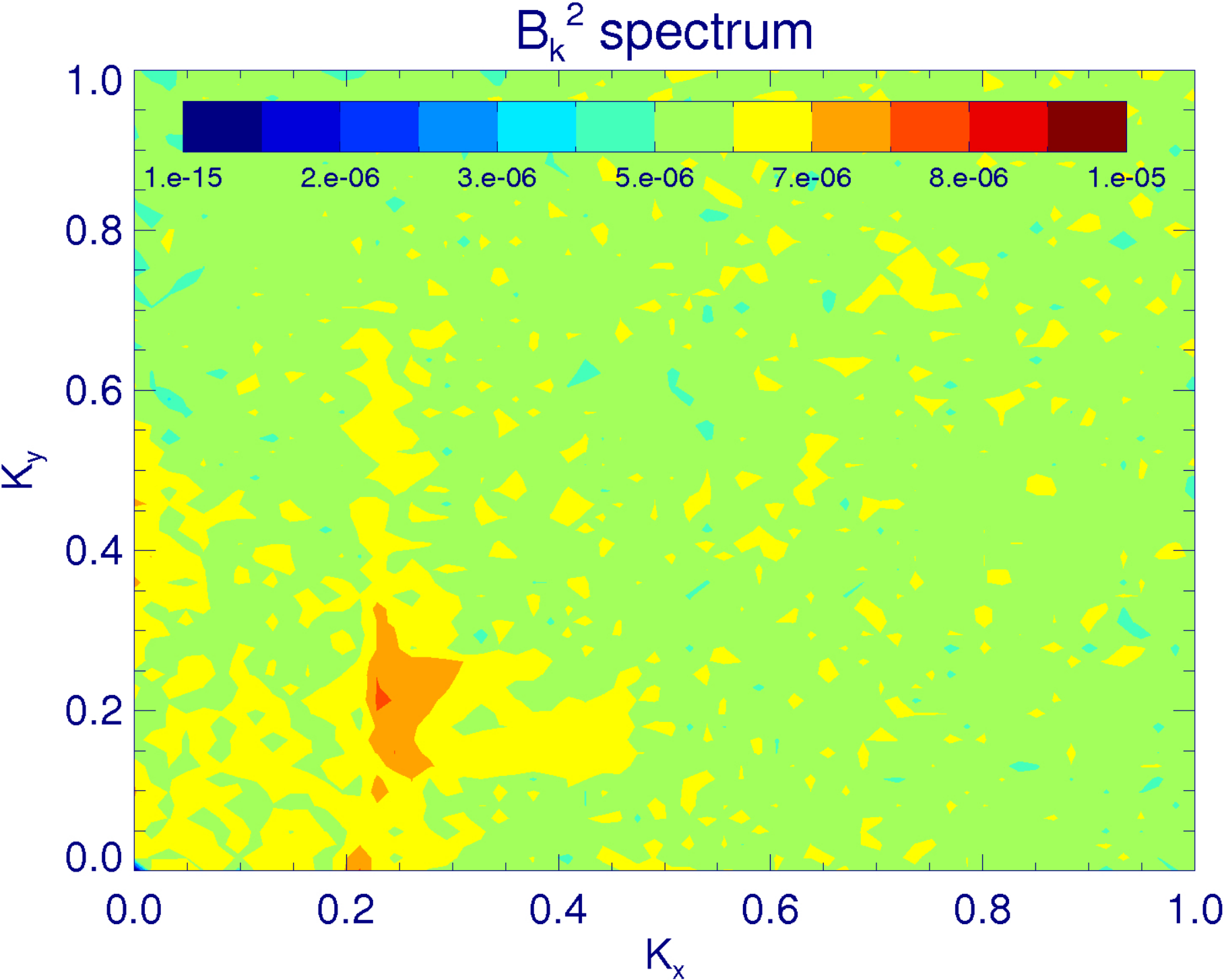}
\caption{Power spectra of the evolved magnetic field fluctuations at the final stage of the simulations at $\Omega_pt= 800$. The figure shows the square of the magnetic field fluctuations as a function of normalized parallel and perpendicular wave-numbers for the case of initially obliquely propagating waves at $\theta=30^\circ$.} 
\label{fig:b2_kx_ky_end_sim_30deg}
\end{figure}

\begin{figure}
\centering
\includegraphics[width=0.45\textwidth, height=0.45\textheight, angle=90]{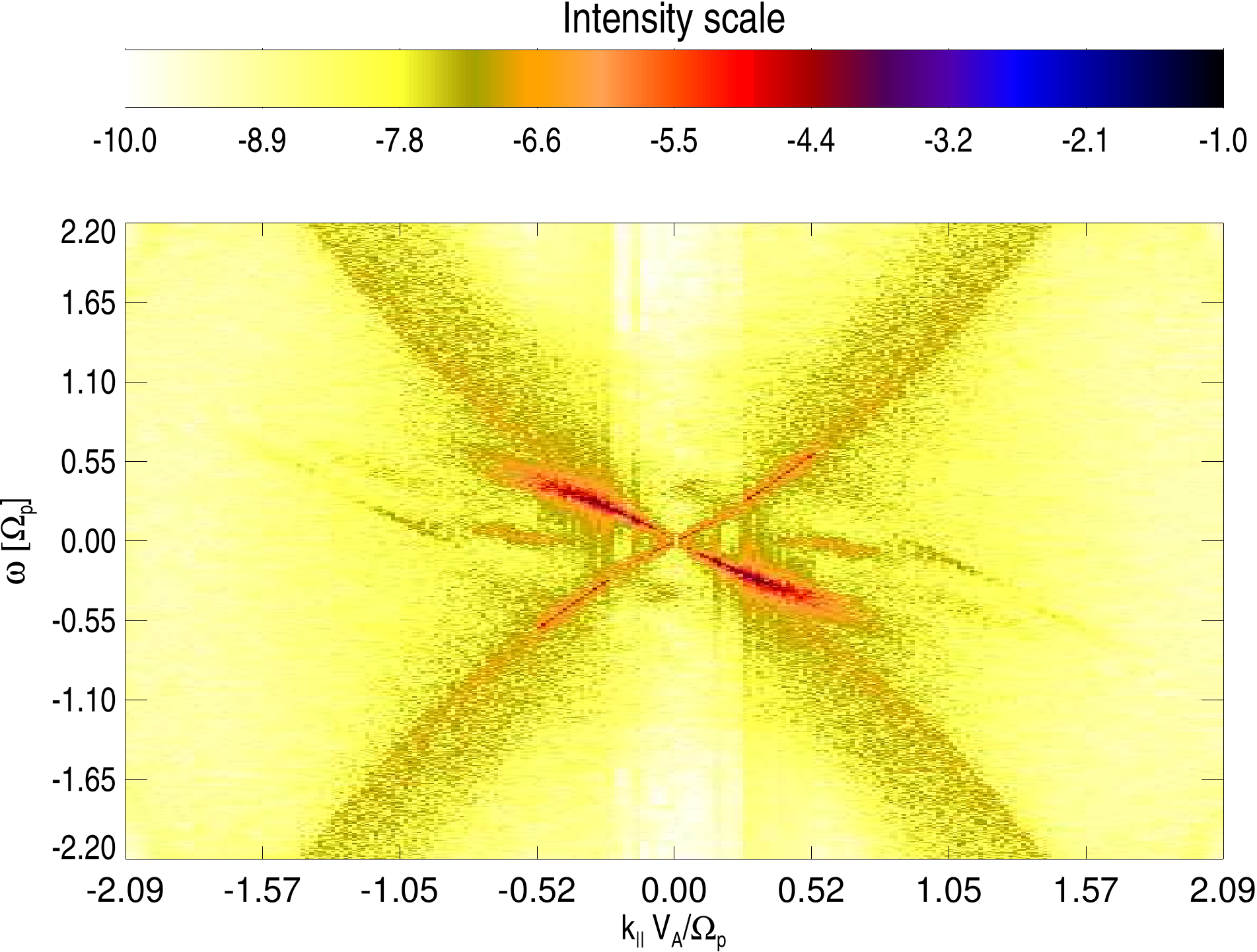}
\hspace{4pt}
\includegraphics[width=0.45\textwidth, height=0.45\textheight, angle=90]{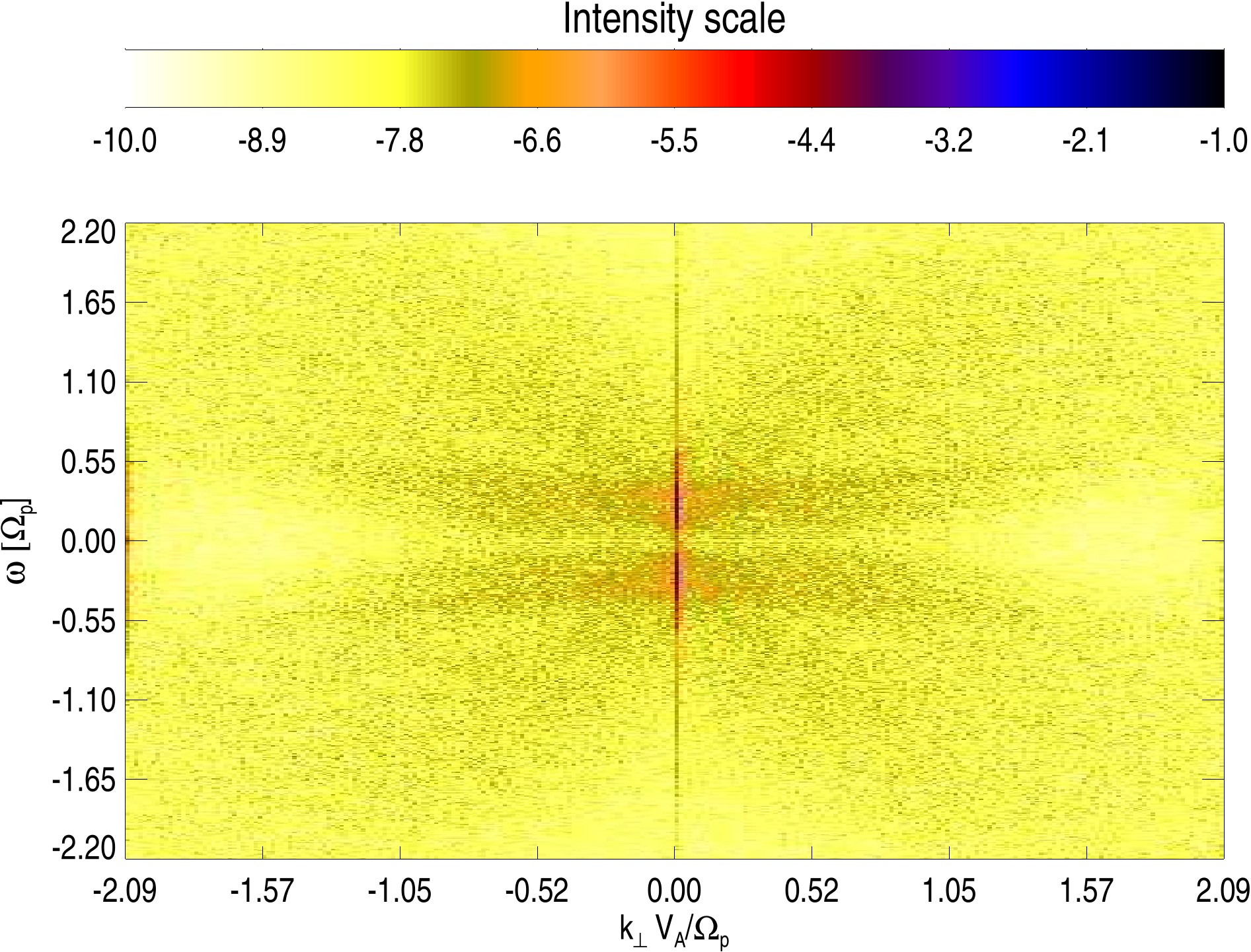}
\caption{Power spectra of the magnetic field fluctuations in Fourier space at the final stage of the simulations
at $\Omega_pt= 800$. The figure shows the square of the magnetic field fluctuations in logarithmic scale as a function of normalized frequency, parallel (top panel) and perpendicular (bottom panel) wave-numbers for the case of initially parallel wave propagation.} 
\label{fig:wk0}
\end{figure}

\begin{figure}
\centering
\includegraphics[width=0.45\textwidth, height=0.45\textheight, angle=90]{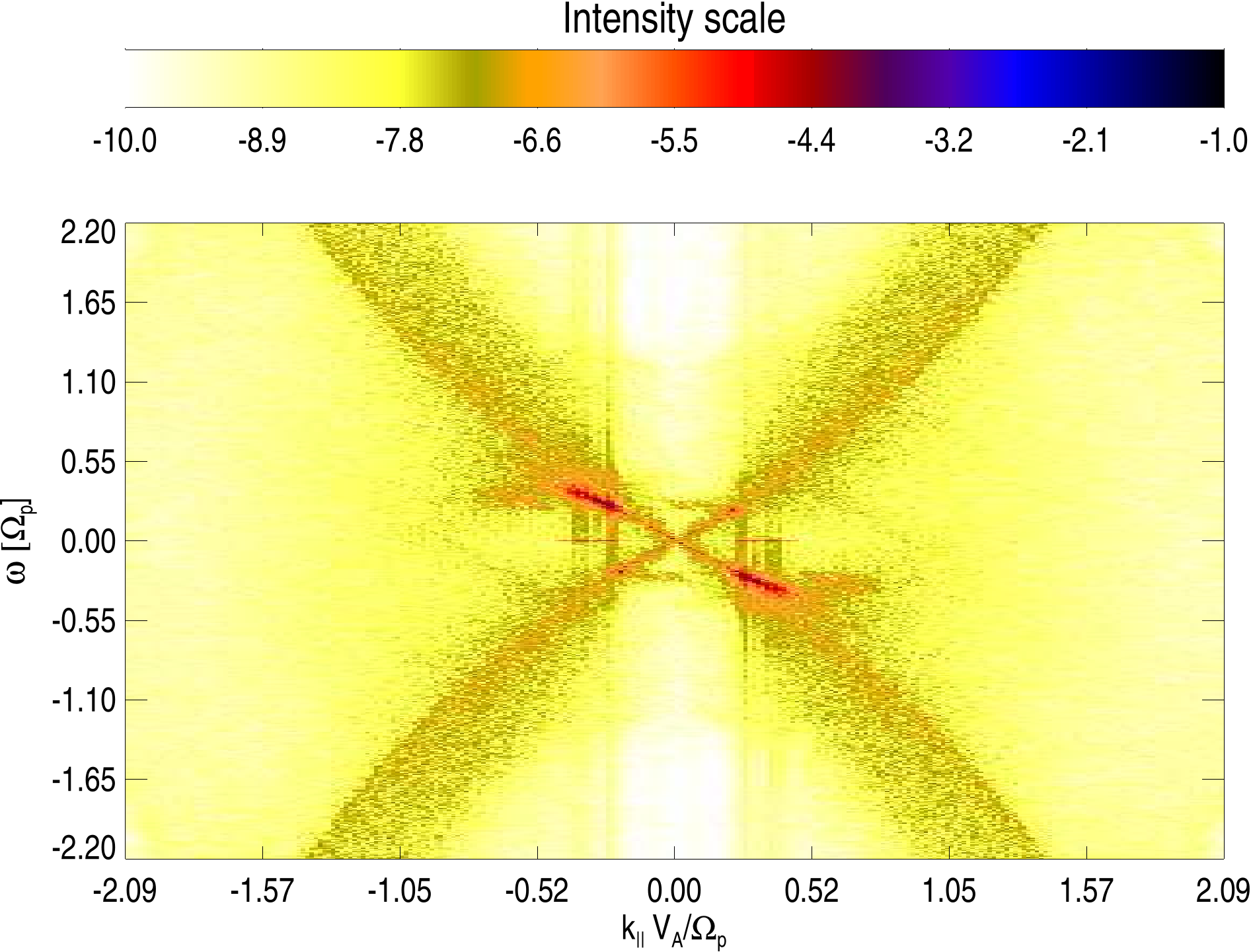}
\includegraphics[width=0.45\textwidth, height=0.45\textheight, angle=90]{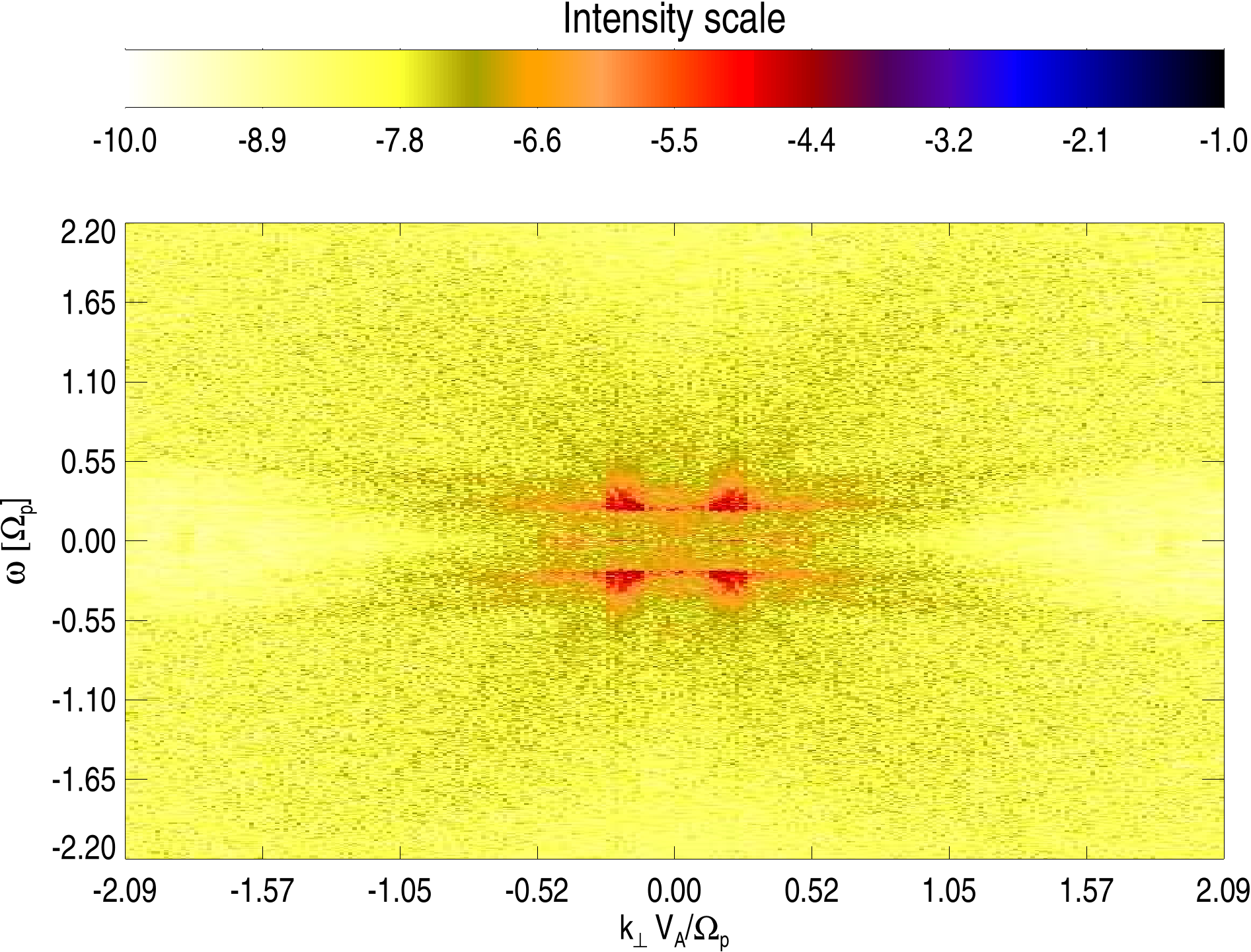}
\caption{Power spectra of the magnetic field fluctuations similar to Fig.~\ref{fig:wk0}, but for the case of oblique waves propagating at $\theta=30^\circ$.} 
\label{fig:wk30}
\end{figure}

\begin{figure}
\centering
\includegraphics[width=0.45\textwidth, height=0.45\textheight, angle=90]{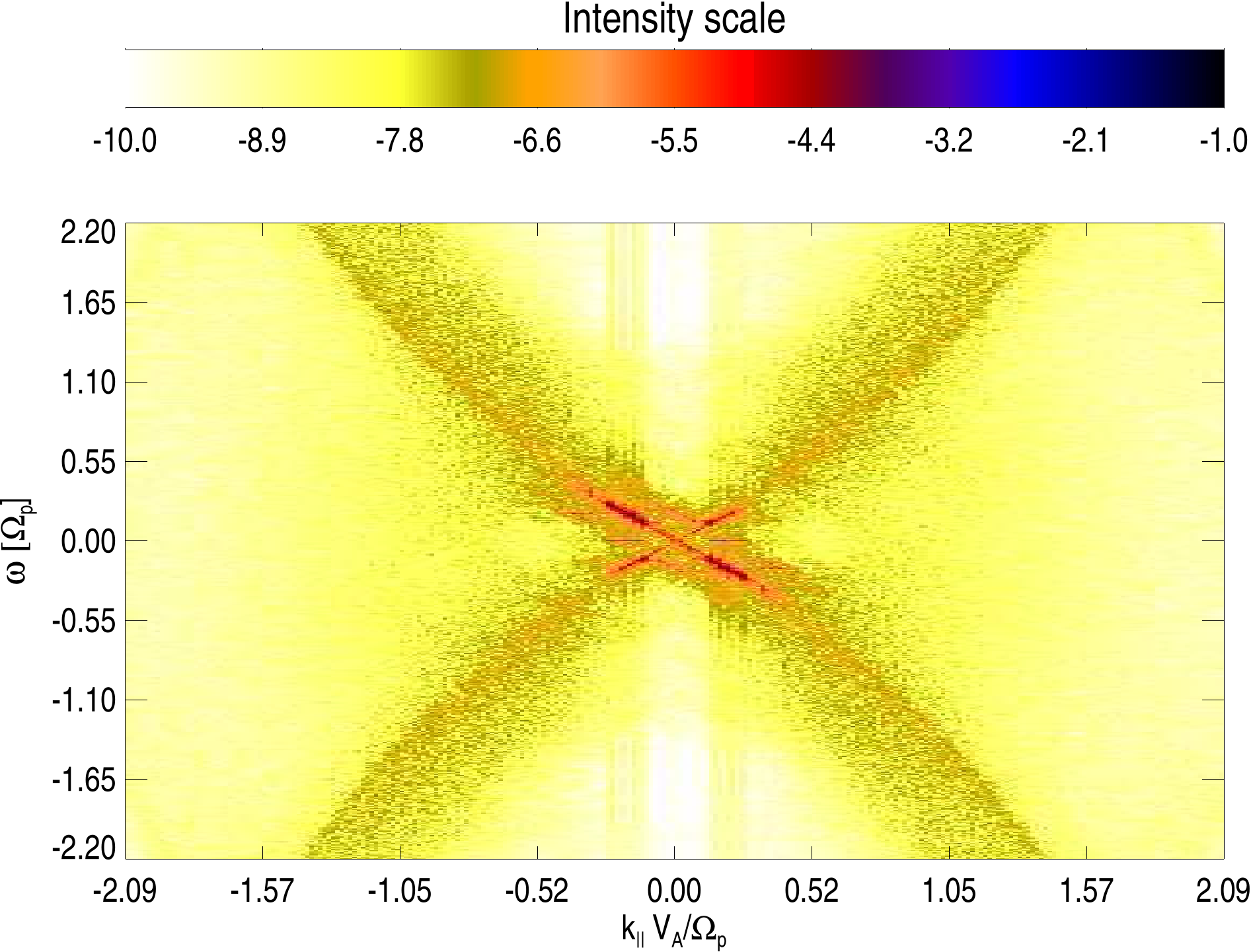}
\includegraphics[width=0.45\textwidth, height=0.45\textheight, angle=90]{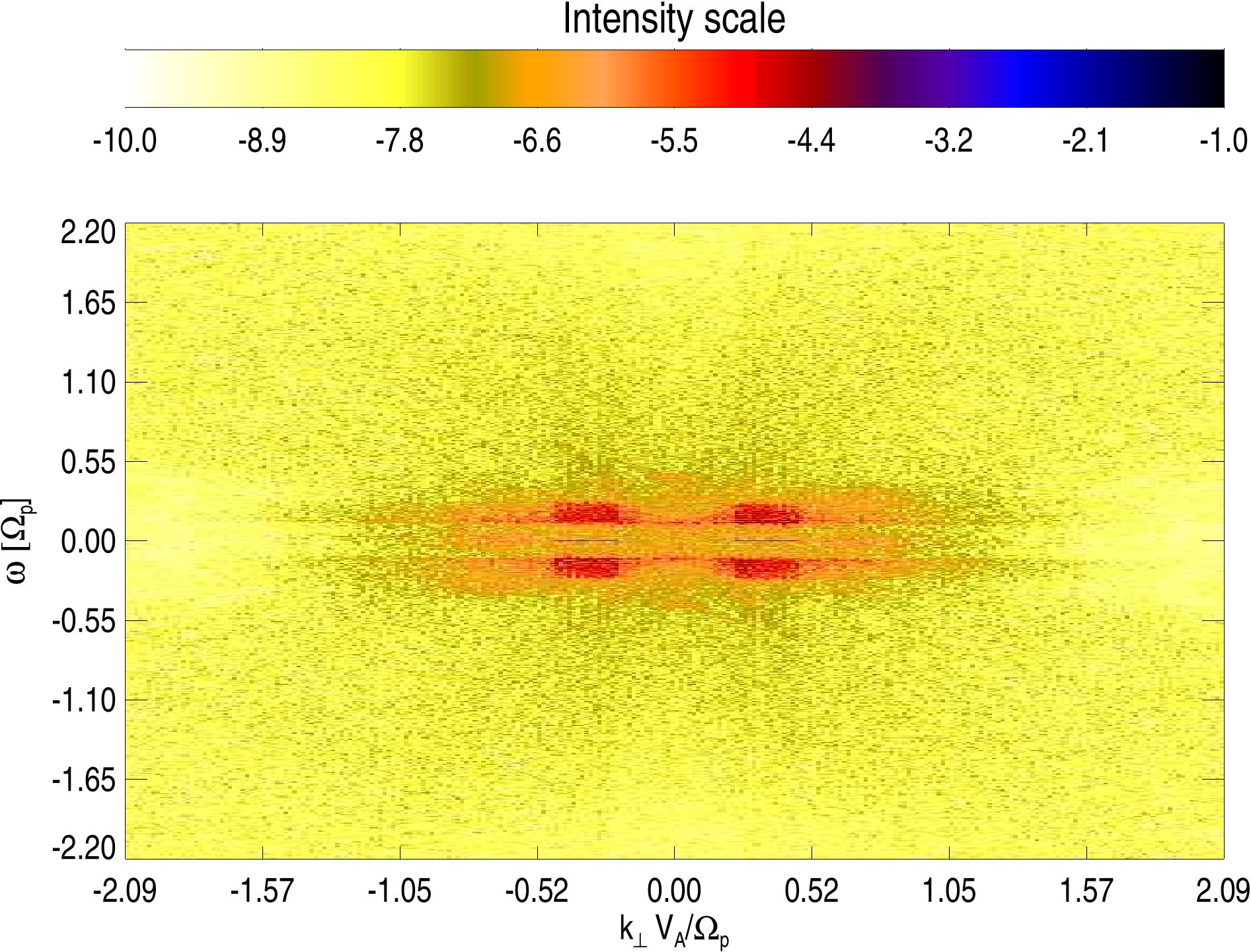}
\caption{The same spectral plot for the magnetic field fluctuations as on Fig.~\ref{fig:wk0} and Fig.~\ref{fig:wk30}, but for the case of initial oblique wave propagation at $\theta=60^\circ$.} 
\label{fig:wk60}
\end{figure}

In this section we will answer the objectives posed in the introduction, namely to compare the relative role of parallel and oblique Alfv\'en-cyclotron waves for the observed preferential heating of minor ions in the fast solar wind near the Earth. As a secondary goal of this research, we will investigate the evolution of the stable point $\tilde{\beta_p}=0.33; T_{\perp p}/T_{\parallel p}=1$ in the central domain of the plasma $\beta$ and temperature anisotropy space for the protons, where the majority of the ion velocity distribution functions are observed \citep{Matteini:07,Bale:09,Maruca:11,Maruca:12,Wicks:13,Hellinger:14}.
This point is stable with respect to linear plasma instabilities and it is interesting to study its temporal evolution in ($\tilde{\beta},T_{\perp p}/T_{\parallel p}$) space in the presence of the ambient solar wind turbulence.
Figure~\ref{fig:anisodr} demonstrates the temporal evolution of the temperature anisotropies for the $\alpha$ particles (top panel), protons (middle panel), as well as the relative drift speed between the two ion species (bottom panel). The initial plasma state represents the \textit{Wind} measurements explained in Sections 2 and 3. The figure presents a comparison between the different simulation cases (see Table~1) with strictly parallel or oblique initial wave spectra with propagation angle $\theta=0^\circ, 30^\circ$ and $60^\circ$, respectively. The initial state for all three cases is isotropic drifting plasma with $V_{\alpha p} =0.44 V_\mathrm{A}.$ The slight initial temperature anisotropy (mainly in the parallel case) is apparent and is due to the non-thermal contribution in the transverse bulk velocity for the ions generated by the initial broad-band wave spectra, see Eq.~(8). The slight initial deviation from the selected value of the relative drift speed in the case of oblique wave propagation is due to the bulk parallel velocity fluctuations induced by the oblique waves.
The temperature anisotropy of the $\alpha$ particles throughout the simulations varies with the propagation angle and is highest for the case of parallel wave propagation with final ${(T_\perp/T_\parallel)}_\alpha \approx 1.24$, whereas the anisotropy of the protons smoothly decreases for all propagation angles considered here. Thus $T_\perp/T_\parallel$ for the minor ions increases with $5\%$ for the case of parallel wave propagation, decreases with $5\%$ for $\theta=30^\circ$ and is reduced by $20\%$ for the highly oblique waves with $\theta=60^\circ$. The anisotropy of the protons is reduced for all cases: it decreases with $21\%$ in the parallel case, $11\%$ in the case of initial waves at $\theta=30^\circ$ and $9\%$ for $\theta=60^\circ$. These changes in the ion temperature anisotropies are related to the preferential heating of the minor ions and the perpendicular cooling for the protons in the decreasing fluctuations of the transverse magnetic field. The relative drift speed slowly decreases within the simulation time of $800$ proton gyro-periods. It decreases with approximately $8\%$ for the case of parallel waves, $15\%$ for the case of oblique waves propagating at $\theta =30^\circ$, and is significantly reduced by $34\%$ for the case of highly oblique wave propagation at $\theta =60^\circ$. We should note that all cases above should be stable from the point of view of linear Vlasov instability.
Figure~\ref{fig:temp} illustrates the temporal evolution of the parallel (along) and the perpendicular (across the ambient magnetic field) components of the ion temperature for protons (top row) and $\alpha$ particles (bottom row) throughout the simulation time. The figure is complementary to Figure~\ref{fig:anisodr} and shows the preferential ion heating for minor ions in the electromagnetic field of the non-resonant initial wave spectra. The plots show that the $\alpha$ particles acquire high perpendicular temperatures for all cases with an estimated growth rate of $65\%$ to $72\%$ as we transition from the parallel to the highly oblique initial wave spectra. The parallel temperatures for the minor ions are also significantly enhanced and their increment throughout the simulations varies from approximately $50\%$ for the initial parallel waves through $70\%$ for the slightly oblique case and reaches as high as $120\%$ for the case of highly oblique initial wave spectrum at $\theta=60^\circ$. The dominant parallel heating in the latter case leads to parallel temperature anisotropies for the $\alpha$ particles with $T_\perp/T_\parallel <1.$ The intense minor ion heating in both parallel and perpendicular directions can be attributed to non-resonant particle scattering and absorption of the initial low-frequency transverse fluctuations as well as to Landau and cyclotron damping of resonant $\alpha$-cyclotron waves generated via nonlinear cascade in the course of evolution. The most prominent parallel heating of the minor ions in the case of highly oblique waves is due to parallel electric field associated with the electromagnetic Alfv\'en-cyclotron waves at oblique wave propagation. In the strictly parallel case only longitudinal waves can carry electric field fluctuations and the initial spectra of transverse Alfv\'en-cyclotron waves cannot contribute to the electric field and density fluctuations, unless nonlinear wave-wave interactions, such as parametric instabilities or more complicated wave couplings take place. Ion heating by parametrically unstable Alfv\'en-cyclotron waves has been discussed for example in \citep{Araneda:09} and \citep{Maneva:14}.
The protons are less affected by the low-frequency initial wave-spectra than the $\alpha$ particles. Their parallel temperature throughout the simulations time increases within $5-8\%$ and their perpendicular temperature decreases adiabatically by $10-14\%$ with the decay of the initial transverse magnetic spectra.
%
%
%
Figure~\ref{fig:vdf_p0_30_60} and Figure~\ref{fig:vdf_a0_30_60} describe the final stage (at $\Omega_pt=800$) ion velocity distribution functions for protons and $\alpha$ particles for the three different angles of propagation of the initial Alfv\'en-cyclotron wave spectra. The left panel shows the case of initially strictly parallel waves, the middle panel describes oblique propagation at $\theta=30^\circ$ and the right panel shows the evolved velocity distributions for the highly oblique case with $\theta=60^\circ$. The bottom rows show the particle counts along the parallel direction (solid lines), over-plotted with the best bi-Maxwellian fit (dotted lines). The center of all velocity distribution functions is shifted due to the presence of relative drift between the $\alpha$ particles and protons. 
The figures indicate that the final distributions for both ion species deviate from the drifting bi-Maxwellian approximation. Due to the perpendicular cooling and the parallel heating the velocity distributions for the proton species are elongated at all propagation angles with a distinct onset of a beam formation in the parallel case. The scattering at the front side (the right hand side) of the distribution becomes less prominent in the case of highly oblique wave propagation and at $\theta=60^\circ$ the proton distributions can be sufficiently well-described within the drifting bi-Maxwellian approximation.
The velocity distribution functions for the $\alpha$ particles show strong signatures of particle scattering at the front part of the distributions for all propagation angles and at the tail of the distribution functions in the cases of oblique wave propagation at $\theta=30^\circ$ and $\theta=60^\circ$. This scattering is caused by the wave-particle interactions with the turbulent spectra and relates to the observed perpendicular and parallel heating for the minor ions. The distributions are skewed and throughout the simulation time we observe prominent forward beam formation for both protons and the $\alpha$ particles in the case of initial parallel wave spectra and a combination of forward and inverse $\alpha$ beam formations (referring to the positive/front and negative/back side of the velocity distribution function) in the case of oblique wave propagation at $\theta=30^\circ$ and $\theta=60^\circ$. In the highly oblique case the back scattering by the oblique waves dominates and a dominant inverse $\alpha$ beam is formed. Similar beam formations through particle scattering by initial parallel wave spectra and monochromatic waves have been reported earlier in previous hybrid simulation studies, for example \citep{Maneva:13a}, \citep{Maneva:14} and \citep{Maneva:15}.
Figure~\ref{fig:bx_x_y_end_sim} shows the parallel component of the normalized magnetic field at the final stage of the simulation, $\Omega_pt=800$, for the three different initial propagation angles. The scale is given in units of the magnitude of the constant magnetic background, whose contribution has not been subtracted from the fluctuations. The top panel shows the generated magnetic filed fluctuations in parallel direction for the case of initially parallel wave propagation with $\theta=0^\circ$. The initial strictly parallel transverse wave spectra includes no variations along the direction of the background field and the observed small level of fluctuations at the final stage are entirely due to the generation of oblique waves. In the case of oblique wave propagation at $\theta=30^\circ$ there is a structured redistribution of the initial magnetic field fluctuations in real space, enhancing the wave power in some regions of the box, for instance at $x > 300 d_i$ and $y < 100 d_i$, see Figure~\ref{fig:bx_x_y}. In the case of initial wave propagation at $\theta=60^\circ$ the final stage magnetic field also shows a structured form different from the shape of the initially constructed fluctuations. In both cases in the course of evolution the waves start to form more coherent patterns. The total amplitude of magnetic field fluctuations decreases over time for all three cases.
Figure~\ref{fig:b2_kx_ky_end_sim_30deg} depicts the evolution of the initial magnetic field power in the Fourier space for the case of obliquely propagating initial wave spectra at $\theta=30^\circ$. The figure depicts the strong depletion of the initial wave power with significant direct cascade in perpendicular direction particularly visible around $k_y = k_\perp = 0.36 \Omega_p/V_\mathrm{A}$ and $k_y = k_\perp = 0.46 \Omega_p/V_\mathrm{A}$. We should note that the initial power is transferred to a wide range of perpendicular wave-numbers, which are not initially present in the system (note that the initial range of perpendicular wave numbers at $\theta = 30^\circ$ is $k_\perp \in [0.13, 0.26] \Omega_p/V_\mathrm{A}$). In addition substantial fraction of the wave power has cascaded towards strictly perpendicular wave numbers, which implies the generation of dispersionless wave modes propagating at $\theta = 90^\circ$. As visible in the next figures this process persists for the case of initial spectra at $\theta = 60^\circ$ where the wave energy is transferred to even smaller transverse scales with higher perpendicular wave-numbers. Apart from the perpendicular cascade the spectra also shows both direct and inverse cascade in parallel direction. The direct cascade happens at earlier stage before the end of the simulations and generates resonant modes, which are quickly absorbed by the minor ions.
Figures~\ref{fig:wk0} through \ref{fig:wk60} provide information about the nonlinear evolution of the initial wave-spectra and the dispersion of the turbulence generated waves in the course of the simulations. The simulation dispersion relation is computed from the power of the Fourier transformed magnetic field fluctuations in space and time. To compute the dispersion in parallel direction, the Fourier transformation is taken at a fixed position, from the vertical domain of the initial wave spectra. The same procedure is used in the computation of the wave dispersion in perpendicular direction. To compute the dispersion we have used the temporal information throughout the end of the simulations.
The magnetic field power along the vertical line perpendicular to $k_\perp$ in the bottom panel of Figure~\ref{fig:wk0} is due to the strong parallel waves, plotted in the top panel. The horizontal lines with $\omega < 0.5 \Omega_p$ may relate to ion Bernstein modes for the $\alpha$ particles, while higher-frequency proton-Bernstein modes are highly damped. The figure indicates the generation of strong forward propagating fast modes, as well as strong backward propagating Alfv\'en-cyclotron modes at wave-numbers range spreading beyond the wave-numbers of the initial parallel wave spectrum. Low-frequency backward propagating Alfv\'en waves are undamped solutions of the parallel linear Vlasov dispersion relation when sub-Alfv\'enic drifts are present. Here they are excited through nonlinear wave-wave interactions in the presence of the initial wave spectra and the induced ion temperature anisotropies. The generation of these background propagating waves continues at oblique wave propagation, but they are limited within lower wave-number range and at $\theta=60^\circ$ their signature is overcome by strong backward-propagating fast modes, as visible from the next figures.
The forward propagating fast modes are less prominent for the case of $\theta=30^\circ$ and are overcome by backward propagating fast modes when $\theta=60^\circ$.
The zero-frequency lines at the bottom panels in Figure~\ref{fig:wk30} and Figure~\ref{fig:wk60} describe non-propagating entropy waves, generated by the density and pressure fluctuations. In an isotropic and slightly anisotropic plasma with sub-Alfv\'enic drifts within the specific wave-number range linear instability theory predicts that these waves should be stable or fully damped. The presence of the relative drift speed reduces the damping of the slow modes and the entropy waves and the pressure fluctuations associated with the initial wave spectra further enhance the entropy waves wave-power. Thus the positive growth rate of the entropy waves is a nonlinear effect, which can be attributed to the presence of the initial Alfv\'en-cyclotron wave spectra. Looking at the complementary spectral plot shown on Figure~\ref{fig:b2_kx_ky_end_sim_30deg} and the explanations thereafter, the zero-frequency modes can be attributed to the generation of strictly perpendicular Alfv\'en waves.

\section{Discussion and Conclusions}
Parallel and obliquely propagating ICWs are ubiquitous in the solar wind plasma and have been observed at various heliocentric distances. Furthermore, differentially streaming non-thermal ions with beams in their velocity distribution functions and different temperatures along and across the direction of the ambient magnetic field are common features of the fast solar wind streams. Yet, the correlations between the existing plasma waves and the particle properties in the multi-species solar wind are still poorly understood, and further theoretical and simulation studies directly related to observations are needed. In this paper we present the results from observations-driven 2.5D hybrid simulation study, as a first attempt to distinguish the relative role of parallel versus oblique wave propagation for the generation and evolution of ion temperature anisotropies and relative drifts in the fast solar wind streams. The initial state is based on \textit{Wind} data from 16$^\mathrm{th}$ of January 2008, 15:19:09 UT time in the undisturbed wind at 1AU, and we study the evolution of the expanding solar wind plasma parcel, which cannot be followed by the spacecraft. We have investigated the evolution of a stable region within the velocity distribution functions of proton and $\alpha$ particles in anisotropy-$\beta$ space. The initial isotropic distributions of the minor ion species interact with the wave spectra and acquire slightly higher perpendicular temperatures. In contrast the protons cool down, as there is no wave energy at the proton scales. Our findings show that the minor ions are preferentially heated by the magnetic fluctuations for all initial wave spectra considered here, where protons are prone to perpendicular cooling, as expected by the double-adiabatic expansion model predictions. At the final stage of the simulations the strength of the magnetic fluctuations decreases as the magnetic power is converted into thermal energy for the minor ions.
The main assumption in this work is the type of initial wave spectra considered here. Since we are interested in the properties of the minor ions and try to couple the physics at fluid and ion scales, for the initial broad-band wave spectra we have chosen low-frequency left-hand circularly polarized Alfv\'en-cyclotron waves. We exclude the effect of the high-frequency (much higher than the proton-cyclotron frequency) whistler and KAWs waves, which interact mainly with high-speed electrons and proton beams. Instead our study is focused on the direct cascade of wave power from the intermediate fluid regime down to the kinetic ion scales and neglect the effects of inverse cascade at scales much smaller than the proton skin depth or their possible contributions to the energy transfer to the ion scales. Yet, apart from the direct cascade towards the smaller ion scales, our simulations do show signatures of inverse cascade towards larger fluid scales with generation of strictly perpendicular waves as clear from the magnetic field power spectra plot at the end of the simulations.
%
%
%
In this study we have performed 2.5D hybrid simulations to investigate the minor ion heating by broad-band spectra of oblique Alfv\'en-cyclotron waves and the evolution of an initial relative drift speed between the protons and the minor ions. We have compared the ion heating corresponding to different angles of propagation ($0\degr, 30\degr$ and $60\degr$) and found prominent perpendicular heating for the minor ions at all propagation angles. For the selected spectral range of the initial wave spectra for the simulations we find that the $\alpha$ particles are most prominently heated in both parallel and perpendicular direction by the highly oblique Alfv\'en-cyclotron waves propagating at $\theta=60^\circ$. The parallel heating for both species increases with the propagation angle and the parallel temperature of the minor ions in the highly oblique case prevails over the perpendicular one, so that they acquire parallel temperature anisotropy with $T_\perp/T_\parallel <1$. Since the perpendicular heating of the minor ions remains at similar level for all propagation angles the minor ions have highest anisotropy for the case of initially parallel propagating waves.
For all cases considered here the initially isotropic protons experience moderate cooling in perpendicular direction, and slight heating in parallel direction, the net effect of which leads to the generation of parallel temperature anisotropies. Contrary to the linear instability theory predictions the initial relative drift speed $V_{\alpha p} = 0.44V_\mathrm{A}$ decreases in time throughout the simulations. It remains quasi-conserved in the parallel case and gets further reduced in the oblique cases with a magnitude change of over $30\%$ when $\theta=60\degr$. As part of this study we have also looked at the effect of a gradual solar wind expansion with $\varepsilon t_0= 10^{-5}$. As expected by construction, the expansion leads to perpendicular cooling for both species, however considering the realistically small value of the expansion parameter it does not have significant effect on the ion heating within the simulation times considered here.
The current study is concentrated on warm plasmas with $\tilde{\beta} \approx 0.33$, but this result is expected to hold even stronger when applied to the lower plasma $\beta$ conditions in coronal holes and the fast solar wind near the Sun, where $\tilde {\beta} \sim 10^{-2}$. 
Next, we investigate the power spectra of the magnetic fluctuations. We find limited turbulent cascade in the perpendicular direction for the parallel case, accompanied by a strong inverse cascade in  parallel direction towards smaller wave-numbers. For the oblique cases we find prominent additional oblique mode generation in a wider range of wave-numbers than the initial values, as well as the generation of strictly perpendicular waves. Right-hand polarized fast waves are being generated in all simulation cases with strong forward propagating modes in the parallel case and strong backward-propagating modes in the highly oblique case. Strong forward-propagating ion-cyclotron modes are excited in the parallel case, as they are expected normal modes in drifting two ion-species systems. In the cases of initial oblique wave propagation the initial Alfv\'en-cyclotron wave spectrum is strongly damped and generation of entropy modes/perpendicularly propagating Alfv\'en waves is observed.
We should note that the total magnetic energy is gradually depleted throughout the simulation time for all three initial angles of wave propagation and it decreases approximately by $70\%$ from the initial value. Simultaneously the minor ions are continuously heated on average by $69\%$ in perpendicular direction. This analysis suggests that major part of the minor ion heating is due to the external wave energy sources and can not be simply attributed to redistribution of kinetic energy between the protons and the $\alpha$ particles. Further source for the ion heating can be a nonlinear magnetosonic instability induced by the initial drift-speed in the presence of the wave spectra. The value of the initial relative drift is within the stable regime for linear plasma instabilities, yet it decreases up to $30\%$ in the highly oblique case. This can contribute to the energy source of the initial wave spectra and provide an additional energy source needed for the strong parallel minor ion heating with more than $120\%$ increase of the parallel temperature for the minor ions. Note that the relative decrease of the proton temperature anisotropy, which varies between $13\%$ and $20\%$, is due to a combination of parallel heating and perpendicular cooling. The drop in proton anisotropy is most prominent for the parallel case and weakest for the highly-oblique case, whereas the minor ion heating is strongest for the highly oblique case. This is a clear indication that the observed minor ion heating is related to the initial wave-spectra, rather than energy exchange between the two ion species.\\
Looking into the velocity distribution functions we find strong signatures of wave scattering with prominent forward or inverse minor ion beams being formed by $\Omega_pt=800$. The proton scattering by the oblique waves generates leads to no clear beam formations. This can support the conclusion that stronger proton beams are formed by trapping in the ion-acoustic waves potential generated in the nonlinear evolution of the initial wave-spectra or relate to the fact that it is more difficult to accelerate protons by low-frequency waves in the finite-$\tilde \beta$ plasma considered here in comparison to the lower-$\tilde \beta$ regimes close to the Sun, where the wave amplitudes are also expected to be stronger.
We should note that for the limited wave-number range considered in our study the initial spectral slope of the fluctuations does not play an important role for the nonlinear wave energy transfer and the related ion heating/cooling and acceleration or deceleration. Simulation tests with more gradual initial magnetic field power slope, for instance $\alpha = -1$, show similar results to the ones presented here.
This work does not include post-simulation wave polarization analysis, however based on previous experiments and the presented wave dispersion analysis in Fourier space we believe that right-hand polarized waves are being generated in the course of evolution and the final state consists of a mixture of left- and right-handed waves. In a next improvement of the present analysis we will broaden the spectral range of the initial fluctuations and include different polarization states.
As a future perspective we plan to implement a two-dimensional spectrum of a general type, avoiding the rotation of parallel spectra. We would also consider low-frequency right-hand polarized kinetic Alfv\'en waves and whistlers, and investigate the amount of energy they can deposit to the ions before ultimately dissipating at the electron scales.
\begin{table*}
\tiny
\label{table:SimParam}
\caption{\footnotesize{Initial parameters for the different numerical simulation runs. The table presents the simulation cases as a function of the angle of propagation of the initial Alfv\'en-cyclotron wave spectra, $\theta$, the initial relative drift speed, $V_{\alpha p}$, the range of initial wave frequencies $\omega_0$, parallel, $k_\parallel$, and perpendicular, $k_\perp$, wave-numbers. }}
\centering
  \begin{tabular}{|c|c|c|c|c|c|c|c|c|c|c|}
    \hline
    Case~\# & $\theta$ & $V_{\alpha p} [V_\mathrm{A}]$ & $\omega_0$ [$\Omega_p$] & $k_\parallel$ [$\Omega_p/V_\mathrm{A}$] & $k_\perp$ [$\Omega_p/V_\mathrm{A}$]\\ \hline
    1 & $0^\circ$ & 0.44  & [0.23-0.34] &  [0.26, 0.52] & 0 \\ 
    2 & $30^\circ$ & 0.44  & [0.21-0.32] & [0.23, 0.46] &  [0.13, 0.26] \\ 
    3 & $60^\circ$ & 0.44  & [0.13-0.26] &  [0.13, 0.26]  & [0.23, 0.46]\\ 
   \hline
  \end{tabular}
\end{table*}
%

%
%
%

\begin{acknowledgments}
This work was supported by F+ fellowship at KU Leuven and NASA grant $\#$ NNX10AC56G. We are grateful to L. Ofman for various fruitful discussions and his valuable help in providing the parallel version of the hybrid code. A.~F.~Vi{\~n}as would like to thank the \textit{Wind}/SWE-VIS grant for the support.
\end{acknowledgments}

\bibliographystyle{apj}
\bibliography{Maneva_2dhyb}

\clearpage

\end{document}